\documentclass{elsart}

\usepackage{graphicx}
\usepackage{epsfig}

\usepackage{amssymb}

\begin{document}

%***********************

\def\lsim{\mathrel{\raise3pt\hbox to 8pt{\raise -6pt\hbox{$\sim$}\hss {$<$}}}} 
\newcommand{\be}{\begin{equation}}
\newcommand{\ee}{\end{equation}}
\newcommand{\bea}{\begin{eqnarray}}
\newcommand{\eea}{\end{eqnarray}}
\newcommand{\da}{\dagger}
\newcommand{\dg}[1]{\mbox{${#1}^{\dagger}$}}
\newcommand{\hlf}{\mbox{$1\over2$}}
\newcommand{\lfrac}[2]{\mbox{${#1}\over{#2}$}}
\newcommand{\scsz}[1]{\mbox{\scriptsize ${#1}$}}
\newcommand{\tsz}[1]{\mbox{\tiny ${#1}$}}
\newcommand{\doref}{\bf (*** REF.??? ***)}

%***************

\begin{frontmatter}



\title{The energy transfer process \\ in planetary flybys}


\author[jpl]{John D. Anderson,\corauthref{cor}\thanksref{global}} 
\corauth[cor]{Corresponding author.}
\ead{John.D.Anderson@gaerospace.com}
\author[jpl]{James K. Campbell}
\ead{jkcepc@yahoo.com}
\thanks[global]{Present address: Global Aerospace Corporation, 711 W. Woodbury Road Suite H, Altadena, Ca. 91001}
\author[lanl]{Michael Martin Nieto}
\ead{mmn@lanl.gov}

\address[jpl]{Jet Propulsion Laboratory, California Institute of  Technology,\\
Pasadena, CA 91109, U.S.A.}
\address[lanl]{Theoretical Division (MS-B285), Los Alamos National Laboratory,\\
Los Alamos, New Mexico 87545, U.S.A.}

\begin{abstract}
We illustrate the energy transfer during planetary flybys as a function of time using a number of flight mission examples.  The energy transfer process is rather more complicated than a monotonic increase (or decrease) of energy with time.  It exhibits temporary maxima and minima with time which then partially moderate before the asymptotic condition is obtained.  The energy transfer to angular momentum is exhibited by an approximate Jacobi constant for the system. 
We demonstrate this with flybys that have shown unexplained behaviors: i) the possible onset of the ``Pioneer anomaly" with the gravity assist of Pioneer 11 by Saturn to hyperbolic orbit (as well as the Pioneer 10 hyperbolic gravity assist by Jupiter) and ii) the Earth flyby anomalies of small increases in energy {\it in the geocentric system} (Galileo-I, NEAR, and Rosetta, in addition discussing the Cassini and Messenger flybys).  Perhaps some small, as yet unrecognized effect in the energy-transfer process can shed light on these anomalies.  
\end{abstract}

\begin{keyword}
planetary gravity assist, dynamical anomaly
\PACS 
95.10.Eg,  %  (Orbit determination and improvement), \\
95.55.Pe,  % (Lunar, planetary, and deep-space probes),   \\
96.12.De   % (Orbital and rotational dynamics)
\end{keyword}
\end{frontmatter}

%**************************************************

%********************************** 1

\section{Planetary flybys}
\label{intro}

The use of planetary flybys for gravity assists of spacecraft became of wide interest during the 1960's, when the Jet Propulsion Laboratory (JPL) first started thinking about what became the ``Grand Tours" of the 1970's and 1980's (the Voyager missions).  The concept was to use flybys of the major planets to both modify the direction of the spacecraft and also to add to its heliocentric velocity.  
At the time many found it surprising that energy could be transferred to a spacecraft from the orbital-motion angular-momentum of a planet about the Sun, despite the fact it had been known since the works of Lagrange, Jacobi, and Tisserand on the three-body problem 
\cite{moulton, danby}, that the energies of comets could be affected by passing near Jupiter.\footnote{In that same period observations that the solar system could evolve \cite{hills} were similarly met with skepticism.}    

Even in the simplest, circular restricted 3-body problem \cite{danby}, it is {\it not} that the energy of each object is conserved, only that the total energy of the entire system is conserved.
Flybys can both give kinetic energy to a spacecraft (to boost its orbital velocity) and also can take kinetic energy from it (to slow it down for an inner body encounter).  

Hohmann developed a powerful analysis tool for gravity-assist navigation, the method of patched conics \cite{hohmann,wiesel}.  At JPL clearer understandings of gravity assists were obtained from the works of \cite{min} and  \cite{flandroref}.    
In this technique, a two-body (Kepler) orbit about the Sun is first kinematically patched onto a two-body hyperbolic orbit about the planet.  Then the spacecraft proceeds along the planet-centered hyperbolic orbit during the flyby, and finally a new two-body orbit about the Sun is patched onto the post-flyby hyperbolic orbit. 

The question arises in such an approach as to where to do the patch; obviously somewhere along the hyperbolic asymptotes, but where? This can be addressed by considering how large a region of space is controlled by the flyby planet. When the region is reduced to a planet-centered sphere of radius $r$, and the distance between the Sun and planet is given by $R$, a useful {\it sphere of influence} is given by the {\it Roche limit}, $r/R = (3m/M_\odot)^{1/3}$, where $m/M_\odot$ represents the ratio of the planet's mass to the Sun's mass. However, the point where the errors in the respective two-body orbits about the Sun and planet are equal is determined not by the exponent 1/3, but by the exponent 2/5 as derived by Lagrange \cite{wiesel}. 
It actually makes little difference which exponent is used to define this matching point, as this whole calculation is an approximation to the actual dynamics and, in practice, the gravity assist trajectory used for space navigation is computed by numerical integration of the complete equations of motion.  

\begin{figure}[h!]
 \begin{center}
\noindent    
\psfig{figure=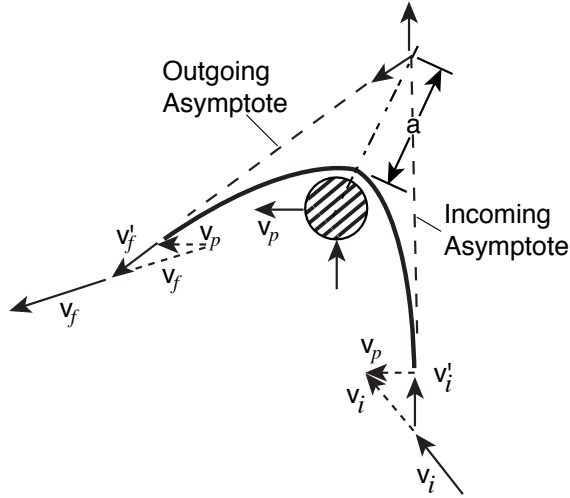,width=3in}  %,height=90mm}
\end{center}
\caption{  \small{Geometry of a flyby, modified from \cite{flandroref}.}   
\label{flandrofig}}
\end{figure}

However, the patched-conic technique is useful for purposes of searching for a fuel-conserving gravity-assist trajectory from a large family of possibilities, and especially when multi flybys of more than one planet are used, for example in the Galileo mission \cite{russell}. In \cite{flandroref} and elsewhere \cite{jdaEGA, VA} the process has been simply and intuitively described and we use it here to gain insights into the nature of the flyby anomaly. 
(See Figure \ref{flandrofig}.)

%****************** Fig. 1

%*****************************

The simple vector velocities in the heliocentric system are added to the orbital velocity of the planet (taken to be constant).  The initial and final velocities in the heliocentric system are ${\bf v}_i$ and ${\bf v}_f$.    The initial and final velocities in the planetary system are ${\bf v}'_i$ and ${\bf v}'_f$.  The velocity of the planet in the solar system is ${\bf v}_p$.  The change in kinetic energy per unit mass is 
\begin{equation}
\Delta {\cal K} = ({\bf v}_f \cdot {\bf v}_f - {\bf v}_i\cdot {\bf v}_i)/2.  \label{fly1}
\end{equation}
A little algebra \cite{flandroref,VA} gives one
\begin{equation}
\Delta {\cal K} = {\bf v}_p \cdot ({\bf v}_f' - {\bf v}_i').                 \label{fly2}
\end{equation} 

Roughly speaking, in the planetary system which rotates anticlockwise, if a satellite in the ecliptic comes from inside the planetary orbit, travels behind the planet, and then goes around it counter-clockwise, kinetic energy will be added to the spacecraft.  Contrarily if a satellite comes from inside the planetary orbit, travels in front of the planet, and then goes around it clockwise, kinetic energy will be taken away from the spacecraft. 

Of course one is not violating conservation of energy.  The energy (and angular-momentum change) is absorbed by the planet that is  being flown by.  However, for such a massive body the relatively infinitesimal change is not noticeable.  Further, there {\it is} (in high approximation) a conserved quantity for the spacecraft in the barycentric system, Jacobi's integral \cite{moulton,danby}:
\be
J = - C/2 = ({\cal K} + {\cal V}) + {\cal L} 
          = ({\cal E}) - \mathbf{\omega \hat{z} \cdot r \times v}, \label{jacobi}
\ee
where $\{{\cal V},{\cal L},{\cal E}\}$ are the potential, rotational-potential, and total energies, respectively, per unit mass, $\mathbf{\omega}$ is the angular velocity of the planet (system) whose vector is aligned with $\mathbf{\hat{z}}$, the unit vector normal to planet's rotational plane.  
Eq. (\ref{jacobi}) is exactly a constant  in the circular restricted 3-body problem, and shows how kinetic energy can be exchanged with angular momentum during a flyby.  

In this paper we are going to discuss two unusual results that are associated with planetary flybys.  The first is the possible ``onset" of the Pioneer anomaly when a major planet gravity assist was used to reach hyperbolic orbit.   The second result is even more surprising, that during Earth flybys at least three craft have exhibited a small velocity increase so that the outbound hyperbolic orbits {\it in the Earth system} have different energies than the inbound orbits.  

In the following sections we will discuss these anomalies from the energy transfer viewpoint.   For the orbits we discuss we obtained information from JPL's Solar System Dynamics web 
site.\footnote{{\tt http://ssd.jpl.nasa.gov/} \label{JPLdynamics}}   
Note that, although this information is precise enough to demonstrate the general energy-transfer properties, to glean out the small anomalies we are ultimately interested in understanding, more precise orbit determinations are necessary, particularly with regard to the Pioneer anomaly.   We  conclude with a discussion.  (In Appendix A we describe how the original trajectory data was obtained and in Appendix B we give further details on the accuracy of the calculations used in this paper.)

% *************************************** 2

\section{``Onset" of the Pioneer anomaly with a flyby boost to hyperbolic orbit?}

Analysis of the radio tracking data from the Pioneer 10/11 spacecraft \\
\cite{pioprl, pioprd}, taken from 3 January 1987 to 22 July 1998
(40 AU to 70.5 AU) for Pioneer 10 and from 5 January 1987 to 1 October 1990 
(22.4 to 31.7 AU) for Pioneer 11, has consistently
indicated the presence of an unmodeled, small, constant, Doppler blue 
shift drift of order $(5.99 \pm 0.01) \times 10^{-9}$ Hz/s. 
After accounting for systematics, 
this drift can be interpreted as a constant acceleration
of $a_P= (8.74 \pm 1.33) \times 10^{-8}$  cm/s$^2$   
directed {\it towards} the Sun, or perhaps as a time acceleration of 
$a_t = (2.92 \pm 0.44)\times 10^{-18}$ s/s$^2$.  This effect has come to be known as the Pioneer anomaly. 
Although it is suspected that there is a systematic 
origin to this anomaly, none has been unambiguously demonstrated.  

\begin{figure}[h!] 
    \noindent
    \begin{center}  
            \epsfig{file=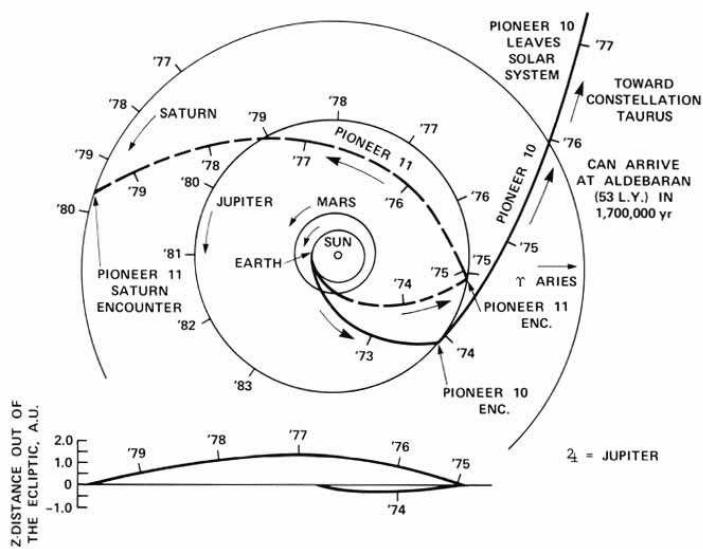,width=3.75in}   %,angle=270}           
\caption{\small{The Pioneer 10 and 11 orbits in the interior of the solar system.}
 \label{earlyorbits}}
    \end{center}
\end{figure}

As to the orbits of the Pioneers (see \cite{edata} for more details), Pioneer 10, launched on 3 March 1972 ET (2 March local time) was the first craft launched into deep space and the first to reach an outer giant planet, Jupiter, on 4 Dec. 1973.  
During its Earth-Jupiter cruise Pioneer 10 was still bound to the
solar system.   With Jupiter encounter, Pioneer 10 
reached solar system escape velocity. 
It is now headed in the general direction opposite to the relative motion
of the solar system in the local interstellar dust cloud.  (Figure \ref{earlyorbits} shows the Pioneer 10 and 11 interior solar system orbits.)

%******************************** Fig 2

%*****************************************

Pioneer 11, launched on 6 April 1973 (ET), cruised 
to Jupiter on an approximate heliocentric elliptical orbit.  
On 2 Dec. 1974 Pioneer 11 reached Jupiter, where it underwent the Jupiter gravity assist that sent it back inside the solar system to catch up with Saturn on the far side.  It was then still on an elliptical orbit, but a more energetic one.  
Pioneer 11 reached Saturn on 1 Sept. 1979.  
After its encounter with Saturn, Pioneer 11 was on an escape hyperbolic orbit.
The motion of Pioneer 11 is approximately in the direction of the Sun's relative motion in the local interstellar dust cloud (towards the heliopause).  It 
is roughly anti-parallel to the direction of Pioneer 10.  

%************  Fig 3

\begin{figure}[h!]
% \vskip -15pt
 \begin{center}
\noindent    
\psfig{figure=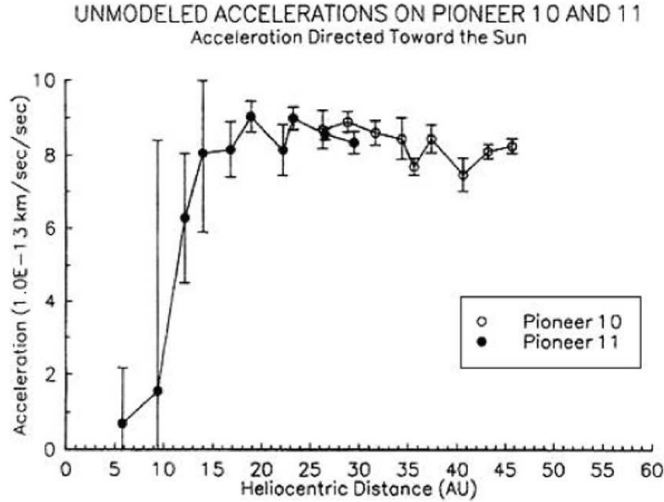,width=3.5in}%,height=110mm}
\end{center}
% \vskip -28pt
  \caption{\small{A JPL Orbit Determination Program (ODP) plot of the early unmodeled 
accelerations of Pioneer 10 and Pioneer 11, from about 
1981 to 1989 and 1977 to 1989, respectively.  (This figure originally appeared in \cite{JPLmemo}.)}
\label{fig:correlation}}
\end{figure} 

%**************

Why this is of interest is that, as early as about 1980 orbit determination results \cite{JPLmemo} began to show an unmodeled acceleration (see Figure \ref{fig:correlation}), what later became known as the Pioneer anomaly \cite{pioprl}  \cite{pioprd}.  
Note that in Figure \ref{fig:correlation} there is an apparent {\it onset} of the anomaly for Pioneer 11 as it passed by Saturn and reached hyperbolic orbit.  
If further study demonstrates that the onset of the Pioneer 11 anomaly was indeed close to the Saturn encounter time and initiation of the unbound hyperbolic orbit, then the obvious question would be if a similar thing happened to Pioneer 10 during its Jupiter encounter, when hyperbolic orbit was obtained \cite{edata}.  

Such further study would be a part of the overall effort to understand the time-dependences and precise direction of the anomaly \cite{edata}.  Since the early trajectory data has been retrieved \cite{stoth}, such work will soon be underway.   In the meantime we here investigate the mechanism of the energy-transfer process.  
In preparation, in 
Table \ref{piotable} we show the orbital elements of the Pioneers 
at the flybys, which resulted in 
hyperbolic escape trajectories from the solar system.

%==================================== 1

\begin{table}[h!]
\begin{center}
\caption{The osculating orbital elements for Pioneer 11 at Saturn closest approach
and Pioneer 10 at Jupiter closest approach.
The quantities are defined in the text except for  
$b$, the impact parameter,   $\Theta$, the 
angle of deflection, $(T,d)$ the time and day of closest approach in UTC, 
and $i$, the inclination.   For Saturn $GM_{S}$ = 37940586 km$^3$/s$^2$
and for Jupiter $GM_{J}$ = 126712766 km$^3$/s$^2$.
\label{piotable}} 
\vskip 20pt
\small{
\begin{tabular}{|l|l|l|}\hline\hline
Quantity &  Pio-11 at Saturn   & Pio-10 at Jupiter      \\   \hline
$ v_{\infty}$ [km/s]    & 7.487  &   8.477   \\
$ v_{F}$ [km/s]        & 31.808  &  36.351    \\
$b$ [km]              & 336,855  & 869,533    \\
$A$ [km]             &   19,062  & 131,279     \\
$\epsilon$            &   1.1176  &  1.1150    \\
$\Theta$ [degrees]   &  126.95    & 127.49    \\  
$i$ [degrees]         &   7.25  &    13.77  \\
$T$        & 16:28:33.370 &  02:25:11.135 \\
$d$        & 01.09.79 &  04.12.73   \\
\hline
\end{tabular} 
}
\end{center} 
\end{table}
%=====================================

%***************************************

%*************************************** 2.1

\subsection{Pioneer 11 at Saturn}

As previously stated, it 
technically is imprecise to consider the total energy of the Pioneer 11 spacecraft separately from that of the rest of the solar system.   Nonetheless, consider 
the 4-body problem with i) the solar-system barycenter (differing from the center because of Jupiter) as the origin of inertial coordinates and with ii) 
the potential energy given by the Sun, Saturn including its leading multipoles (up to octapole), and Titan. 
One can then ask what is the total energy of the spacecraft, $E_P$, with time.    

In the Saturn-centered frame one expects everything to be symmetric before and after encounter. Going to the solar system barycenter frame, to first approximation one {\it a priori} might anticipate a continuous, monotonic transfer of energy with time.  There would be a steady monotonic increase of energy before encounter and then a smooth still-monotonically increasing transition to a new constant energy after encounter. 

\begin{figure}[h!]
\begin{flushleft}
% \hskip 15pt
\noindent
\psfig{figure=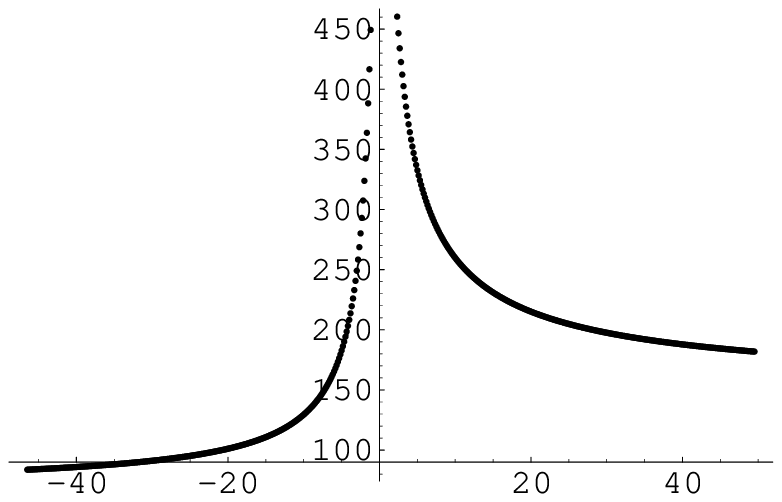,width=2.75in}%,height=75mm}
% \hskip 20pt
\psfig{figure=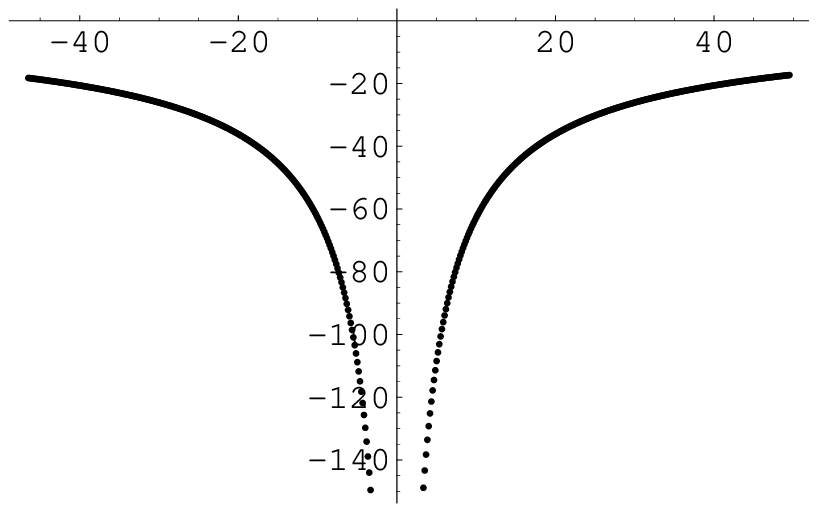,width=2.75in}%,height=75mm}
\end{flushleft} 
\begin{flushleft}
% \hskip 15pt
\noindent
\psfig{figure=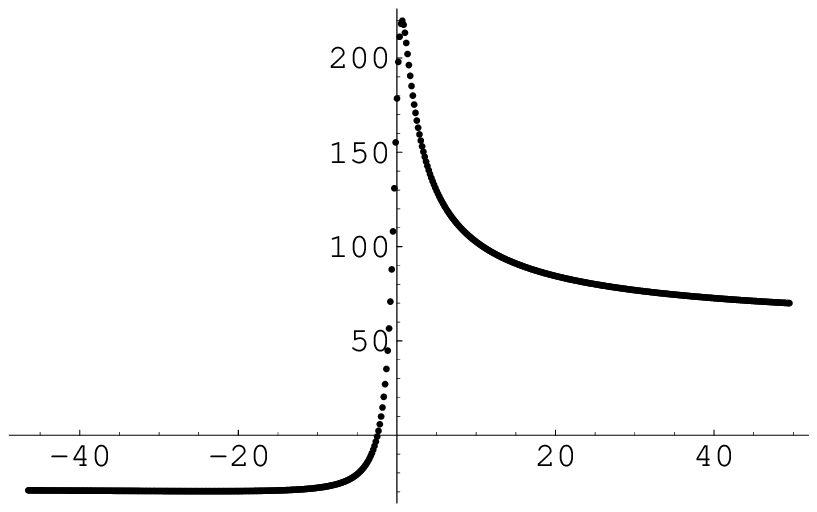,width=2.75in}%,height=75mm}
% \hskip 20pt
\psfig{figure=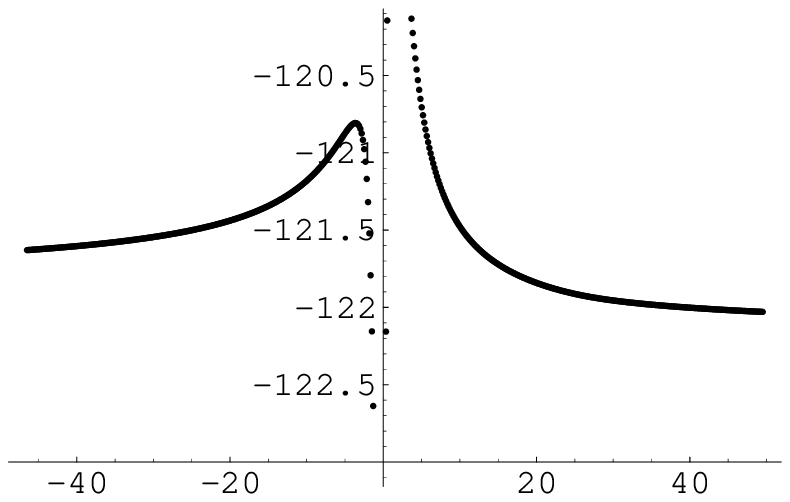,width=2.75in}%,height=75mm}
\end{flushleft} 
\vskip -10pt  
\caption{\small{Energies per unit mass of the Pioneer 11 spacecraft in (km/s)$^2$ plotted vs. time about closest approach to Saturn in hours: the kinetic energy (${\cal K}_P = K_P/m_P$) and Saturn's contribution to the potential energy and next the total energy (${\cal E}_P = E_P/m_P$) and the determined value of the Jacobi ``constant" ($J$) computed using Eq. (\ref{jacobi}) are given.  The latter shows the small residuals about the fit.} 
            \label{11totE}}  
\end{figure}

While at Saturn, the time histories for Pioneer 11's kinetic energy, Saturn's contribution to the potential energy, the total energy, and the determined value of the Jacobi ``constant" (all per unit mass) are given in Figure \ref{11totE}.\footnote{The potential from the Sun varied much more slowly and basically was a bias of $-94.5$ (km/s)$^2$.}

%************ Fig 4

%***********

Figure \ref{11totE} might surprise even though, as we come to below, it was to be expected.  Its up-down asymmetry is huge and there is nothing like left-right symmetry.  Pioneer 11 reached a state of positive total energy about 2-1/2 hours before closest approach to Saturn. But then the spacecraft first gained more than the final energy-shift from the planet and then lost some.  Does this make sense?  If we go back to Section \ref{intro} it does.

Reconsider Figure \ref{flandrofig} and Eqs. (\ref{fly1}) and (\ref{fly2}).  These last are
asymptotic formulae. But we can make them a time-dependent formulae by changing ${\bf v}'_f$ to ${\bf v}'(t)$.  Then, we have 
\begin{equation}
\Delta {\cal K}(t) = {\bf v}_p \cdot ({\bf v}'(t) - {\bf v}_i').
\end{equation}
The second term is a constant.    By looking at Figure \ref{11totE} one can see that the first term gives the spike after perigee with respect to the planet being flown by.  
As the satellite goes around the back side of the planet, 
${\bf v}'(t)$ starts to align with ${\bf v}_p$.  The maximum  of $\Delta {\cal K}$ is reached when ${\bf v}'(t)$ is parallel with ${\bf v}_p$.  This occurs just after perigee.  Then as the satellite swings around further, ${\bf v}'(t)$ goes out of alignment with ${\bf v}_p$ so $\Delta {\cal K}(t)$ decreases some.  The first three graphs of Figure \ref{11totE} now makes sense and it is clear what would happen if the orbit went the other way (gravity-decrease of total energy).

The last graph of Figure \ref{11totE} shows that the energy comes from angular momentum (${\cal L}$).   (See Eq. \ref{jacobi}.)  ${\cal E}(t)$ has been calculated as a function of time and is shown in the third graph of Figure \ref{11totE}.   $\mathbf{\hat{z}}$, 
$\mathbf{r}$, and $\mathbf{v}$ can be calculated.  Therefore,  using $J$, the period $\tau=2\pi/ \omega$, and the declination and right ascension ($\delta_z$ and $\alpha_z$) of the rotation axis $\hat{\mathbf{z}}$ as parameters, a singular value decomposition (SVD) fit  was done  to compare 
\be
{\cal E}(t) = J - {\cal L}(t).     \label{j-fit}
\ee

The three degree of freedom fit results are: 
$J=-121.58$ (km/s)$^2$;
$\tau= 23.97$ yr;
$\delta_z= 54.68^\circ$; and  
$\alpha_z=268.75^\circ$. 
The resulting period differs from Saturn's orbital period of 29.4475 yr by about 19\%, but it represents a best fit to the total energy in the flyby.
The direction of the angular velocity is close to the ecliptic pole at 
$\alpha=270^\circ$, $\delta= 66.5607^\circ$. 

The small deviation of $J$ from a constant is shown in the last graph of  
Figure \ref{11totE}.  
The fact that $J$ is almost the same before and after flyby constitutes Tisserand's criterion for the identification of comets.
What matters is that the absolute values of $J(t)$, which represent a best fit of the fitting model of Eq. (\ref{j-fit}) to the total energy ${\cal E}(t)$ over the flyby interval shown. The time-averaged values of $J$ over the flyby intervals are given in the text. 
Further, the implied plot of ${\cal E}$ is indistinguishable from the third graph in Figure \ref{11totE} on that scale.

$J$ is not exactly constant because the real problem is not derived from a static rotating potential of the circular restricted 3-body problem. The presence of Titan, as well as an oblate Saturn whose rotational axis is not aligned with its inertial orbit, both introduce time explicitly into the rotating system. 
This is demonstrated by the deviations from constancy being closest near periapsis with respect to Saturn.  Note that the Saturn rotational pole is closer in direction to the Earth's pole than it is to the pole of the ecliptic, or the pole of Saturn's orbit. 
 
Here and later note that, since the flyby planet is not strictly in a circular orbit, the circular restricted three-body problem is only an approximation even with no other bodies in the system. The approximation could be made better by introducing the elliptical restricted three-body problem.  However, even with this single improvement the Jacobi integral is no longer a constant of the motion. Instead it has a mean value plus time-varying periodic terms \cite{delva1979}. The constant angular rate $\omega$ in Eq. (\ref{jacobi}) is replaced by a time-varying angular rate given by $\omega(t) = {\mathit l}(t)/r(t)^2$.  (Here ${\mathit l}(t)$ is the orbital angular momentum per unit mass for the planet about the solar-system barycenter and $r(t)$ is the planet's distance from the barycenter.)   To the first order in the eccentricity $e$, $\omega(t) = \omega_0 (1 + 2 e \cos \omega_0 t$), with the time $t$ measured from the flyby planet's perihelion. However, time variations in the Jacobi integral are of concern here only over the time interval of the flyby, which is a small fraction of the total orbital period of the flyby planet. The time variation caused by the planet's orbital eccentricity is a minor contribution when compared to the variations caused by other bodies in the system and by the non-spherical nature of the external gravitational potential of the flyby planet (see Appendix B).

%****************************************************** 2.2

\subsection{Pioneer 10 at Jupiter}

\begin{figure}[h!]
\begin{flushleft}
% \hskip 15pt
\noindent
\psfig{figure=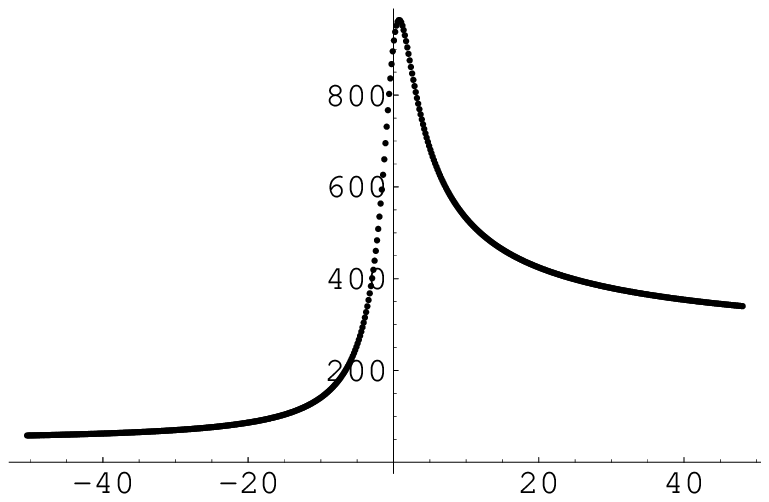,width=2.75in}%,height=75mm}
% \hskip 20pt
\psfig{figure=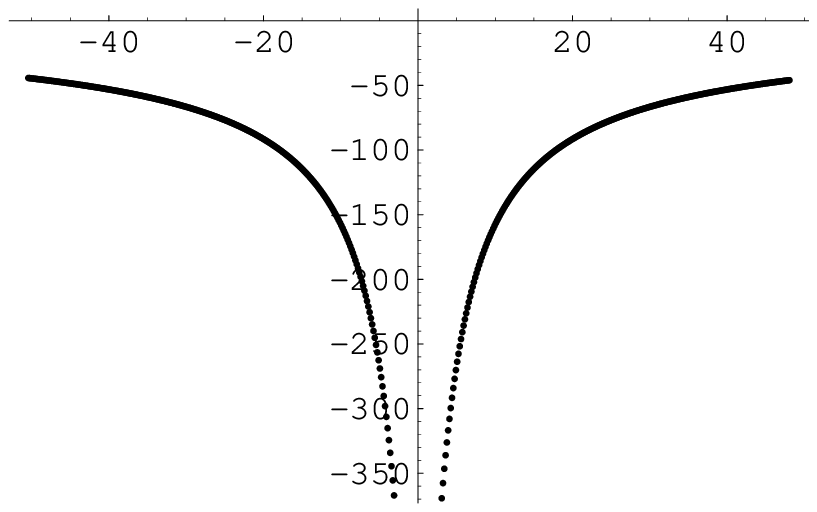,width=2.75in}%,height=75mm}
\end{flushleft} 
\begin{flushleft}
% \hskip 15pt
\noindent
\psfig{figure=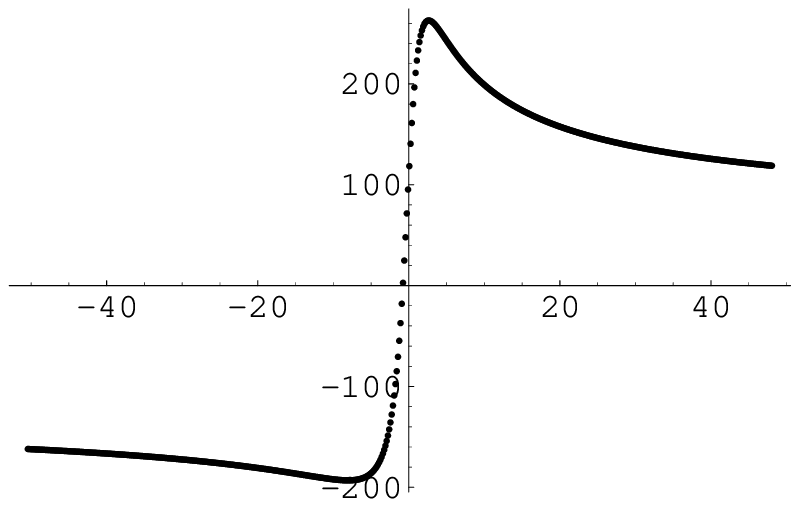,width=2.75in}%,height=75mm}
% \hskip 20pt
\psfig{figure=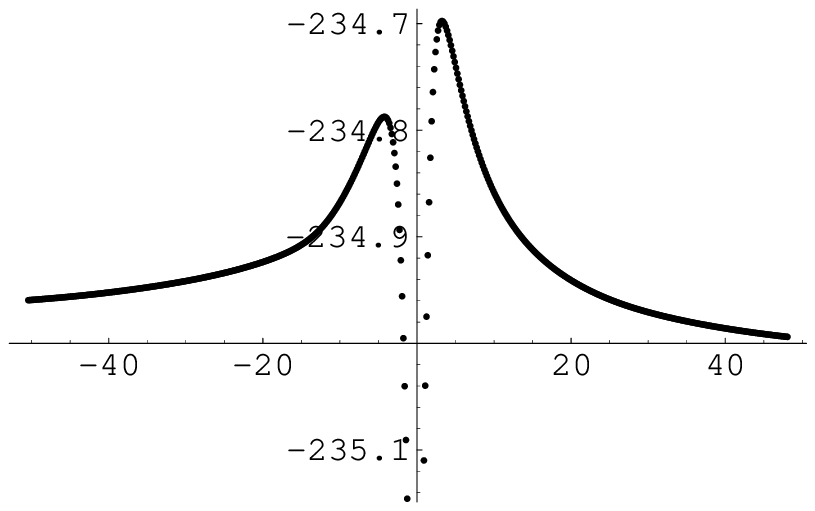,width=2.75in}%,height=75mm}
\end{flushleft} 
\vskip -10pt  
\caption{\small{
Energies per unit mass of the Pioneer 10 spacecraft in (km/s)$^2$ plotted vs. time about closest approach to Jupiter in hours:  the kinetic energy (${\cal K}_P = K_P/m_P$) and Jupiter's contribution to the potential energy and next the total energy (${\cal E}_P = E_P/m_P$) and the determined value of the Jacobi ``constant" ($J$) computed using Eq. (\ref{jacobi}) are given.  The latter shows the small residuals about the fit.
} 
      \label{10totE}}  
\end{figure}

Similarly, in Figure \ref{10totE} we describe the energy-transfer process of the Pioneer 10 flyby of Jupiter, with the solar-system barycenter as the origin of inertial coordinates.  The  potential energy is given by the Sun, Jupiter with its leading multipoles (up to octapole), and the four Galilean satellites. 

Fig. \ref{10totE} first shows, as a function of time,  
the kinetic and the Jupiter-caused potential energies (per unit mass) of Pioneer 10 during its Jupiter flyby.\footnote{The potential from the Sun varied much more slowly and basically was a bias of $-175.5$ (km/s)$^2$.}    
The third graph in Fig. \ref{10totE} shows the total energy as a function of time.  It is similar to the graph in Figure \ref{11totE} except that the total energy exhibits a slight dip before its large rise as it approaches periapsis.

%************ Fig. 5

%***********

Again using $J$, $\tau$, $\delta_z$, and $\alpha_z$ as parameters, an SVD fit using three degrees of freedom was done to compare the two sides of Eq. \ref{j-fit}. 
The results are reasonable:  
$J= -234.93$ (km/s)$^2$;
$\tau=10.70$ yr;
$\delta_z=72.57^\circ$; and
$\alpha_z=297.2^\circ$. 
The resulting period is within 10\% of Jupiter's orbital period of 11.8626 yr as is the direction of the angular velocity with respect to the ecliptic pole direction 
at $\alpha=270^\circ$, 
$\delta=66.5607^\circ$.

One therefore again sees that the energy comes from angular momentum.  Using the values obtained for ${\cal E}(t)$ and ${\cal L}(t)$ yields a ``$J(t)$," given by the last graph in Figure \ref{10totE}. 
Again the Jacobi constant  $J$ is still relatively ``constant."  It is not exactly a constant because the real problem is not derived from a static rotating potential. The presence of the four Galilean satellites and an oblate Jupiter not aligned with its inertial orbit introduces the time explicitly in the rotating system.  
Note that the Jupiter pole is aligned closely to the pole of the ecliptic and to the pole of Jupiter's orbit.

%************************************ 3

\section{The anomalous Earth flybys}

Earth flybys can be an effective technique for increasing or decreasing a spacecraft's heliocentric orbital velocity far beyond the capability of its propulsion system \cite{damar,wiesel}.  It can also be done repeatedly on the time scale of a year.  This technique,
called Earth Gravity Assist (EGA), depends on the bending of the
geocentric trajectory during the flyby, which in turn results in a
change in the direction of the spacecraft's geocentric velocity
vector.    This change in direction can cause either a decrease or increase in {\it heliocentric} orbital energy,
depending on whether the spacecraft encounters Earth on the leading
or trailing side of its orbital path. 

During the flyby the
total energy and angular momentum of the solar system are
conserved.  Further, independent of the heliocentric energy change of the craft itself, 
the spacecraft's total {\it geocentric} orbital
energy per unit mass {\it should} be the same before and after the flyby.  
The data indicates this is not always true.  

Of course, the conservation of the total geocentric energy before and after the flyby is exact in the isolated point mass problem.   However, the total geocentric energy on entering the Earth's {\it sphere of influence} as defined in Section. \ref{intro}, and on leaving it, should be the same, except for terms in the geocentric  gravitational potential that depend explicitly on the time. 
Instead, for Earth flybys by the Galileo, NEAR, and Rosetta spacecraft,
the geocentric orbital energies after the closest approach to Earth
were noticeably greater than the orbital energies before closest
approach.  Further, the changes were much too large for any conventional time-explicit cause, specifically from the Earth's longitudinal harmonics or the motions of the Moon and Sun.
So far, no mechanism, either external or internal
to the spacecraft, that could produce these observed net changes in
orbital energy has been identified.

Table \ref{haletable} shows a summary of the orbital information for these three flybys and also for the interesting Cassini and Messenger flybys.  The values in the Table come mainly from primary sources, and agree to three+ significant figures using orbital information from Horizons.  (See footnote \ref{JPLdynamics} and Appendix B.)  Observe that our altitude, $A$, is the altitude at closest approach above a reference sphere with radius given in Appendix B. These are the radii associated with the gravity fields and are not necessarily the measured mean equatorial radii.
The Messenger numbers, and a few others, come primarily from our Horizons calculations. 

%==================================== 2

\begin{table}[h!]
\begin{center}
\caption{Orbital and anomalous dynamical parameters of  five Earth flybys, 
specifically the osculating elements at closest approach.
The quantities are defined in the text except for  
$b$, the impact parameter,   $\Theta$, the 
angle of deflection, $(T,d)$ the time and day of closest approach in UTC, $(\alpha,\delta)$ the right ascension and declination of the Moon in degrees at closest approach, and $i$, the inclination.   $GM_{\oplus}$ 
is 398600.4415 km$^3$/s$^2$.  The Earth radius, $R_{\oplus}$,
and mass, $M_{\oplus}$, are taken as 6,378.1363 km and $5.9636 \times 10^{24}$ kg.  
\label{haletable}} 
\vskip 20pt
\small{
\begin{tabular}{|l|l|l|l|l|l|}\hline\hline
Quantity &  Galileo (GEGA1)   & NEAR  &    Cassini &  Rosetta & Messenger\\   \hline
$ v_{\infty}$ [km/s]    & 8.949 & 6.851 & 16.01 &  3.863  &  4.056   \\
$ v_{F}$ [km/s]   & 13.738 &  12.739 & 19.03    &  10.517  &  10.389   \\
$b$ [km] & 11,261 &  12,850 &  8,973 &  22,680.49  &   22,319  \\
$A$ [km]      &  956.053  & 532.485   & 1,171.505  & 1954.303 & 2,336.059  \\
$\epsilon$    &  2.4736 &  1.8137 & 5.8564    & 1.312005   &  1.13597   \\
$\Theta$ [degrees]  & 47.67 & 66.92  & 19.66 &  99.396  & 94.7  \\  
$i$ [degrees] & 142.9  &  108.0 &  25.4  & 144.9 &  133.1 \\
$T$ & 20:34:34 & 07:22:56  & 03:28:26   & 22:09:14.12  & 19:13:08  \\
$d$ & 08.12.90 & 23.01.98 & 18.08.99 &  04.03.05  & 02.08.05  \\
$m$ [kg] & 2,497.1  & 730.40  & 4,612.1  &  2,895.2  &   1,085.6  \\
$\alpha$ [degrees] &  163.7  & 240.0  & 223.7  &  269.894  &  107.0\\
$\delta$ [degrees] & 2.975  & -15.37  & -11.16 &  -28.185  &  27.59\\
\hline
$\Delta v_{\infty}$ [mm/s]  & 3.92$\pm$0.08 & 13.46$\pm$0.13 & 
         $\mathbf{\cdots}$   & {1.82$\pm$0.05} &  $\mathbf{\cdots}$\\
$\Delta v_{F}$ [mm/s]   & 2.56$\pm$0.05 & 7.24$\pm$0.07 & 
           $\lsim|-0.2|$   & { 0.67$\pm$ 0.02} &  $\cal{O}$(0)  \\
$\Delta {\cal E}$ [J/kg] & 35.1$\pm$0.7 & 92.2$\pm$0.9 &  
      $\mathbf{\cdots}$   &   7.03$\pm$0.19 &  $\mathbf{\cdots}$\\ 
$\Delta E$ [J] & 87,600$\pm$1,700 & 67,350$\pm$650 &
       $\mathbf{\cdots}$  &   20,350$\pm$560 & $\mathbf{\cdots}$ \\
\hline 
\end{tabular} 
}
\end{center} 

\end{table}
%=====================================

% ********************************** 3.1

\subsection{The first Galileo flyby, GEGA1}

The Galileo spacecraft, launched on 18 October 1989, made use of 
two EGAs to propel it on a trajectory to Jupiter. The first, GEGA1,
occurred on 8 December 1990 at a flyby velocity of $v_F=13.7$ km/s and at an altitude $A=956$ km (flyby radius $r_F = 7,334$ km). 

Soon thereafter, an analysis of radio ranging and
Doppler tracking data revealed that the pre- and post-encounter data
could not be reconciled without the 
introduction of a small but
significant $\Delta v_F \sim 2.5$ mm/s velocity increase at or
near the time of closest approach to Earth \cite{gega1}.  Unfortunately,
during the closest approach the spacecraft was moving too fast and
at too low an altitude for any station of the Deep Space Network
(DSN) to track it.  Hence the time history of the velocity increase
was not obtainable. Even so, a later detailed analysis \cite{ant}
found that, when expressed in terms of the orbital hyperbolic
excess velocity $v_\infty=8.949$ km/s, the increase was
$\Delta v_{\infty} = (3.921 \pm 0.078)$ mm/s. 
The equivalent orbital energy increase per unit mass at flyby was 
\be
\Delta {\cal E}=\Delta E/m_G \approx (v_{\infty})(\Delta v_{\infty}) 
\approx (v_F)(\Delta v_F) 
= (35.09 \pm 0.70) ~ \mathrm{J/kg}.
\ee
(See column 2 of Table \ref{haletable}.)

Now consider GEGA1 from the energy transfer point of view, including the potential energies from the Earth, Sun, the Moon, Jupiter, and the Earth's quadrapole moment.  The last three effects, although interesting, are small.  The Sun's effect is significantly greater, of order 900 (km/s)$^2$, but the change during flyby was small.  (See Section \ref{disc} for comments on the Sun's and Moon's potentials.) That left the Earth's potential energy as the main potential change, this not being surprising of course.

%************ Fig 6

\begin{figure}[h!]
\begin{flushleft}
% \hskip 30pt
\noindent
\psfig{figure=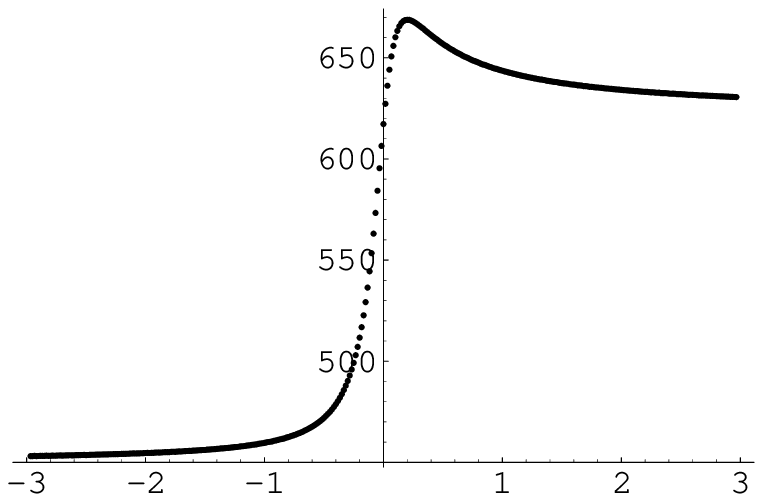,width=2.50in}%,height=75mm}
% \hskip  20pt
\psfig{figure=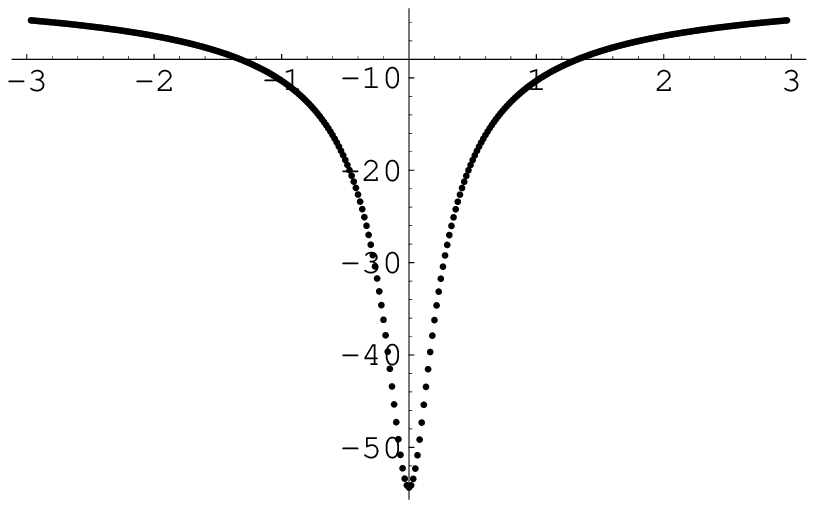,width=2.50in}%,height=75mm}
\end{flushleft} 
\begin{flushleft}
% \hskip 30pt
\noindent
\psfig{figure=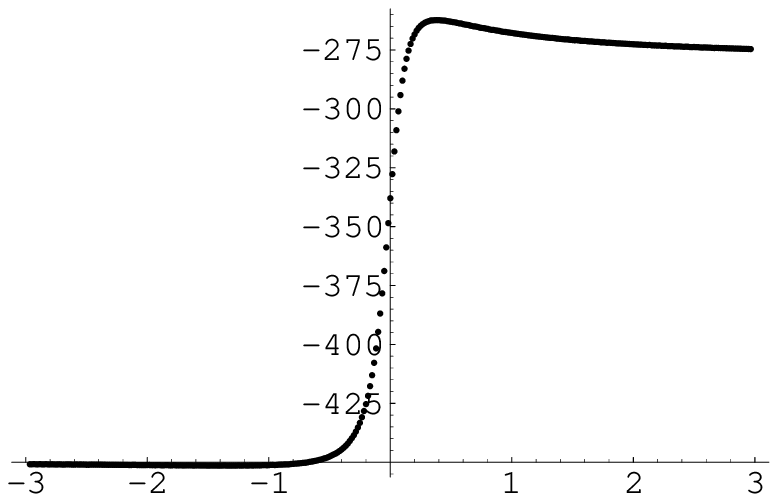,width=2.5in}%,height=75mm}
% \hskip 20pt
\psfig{figure=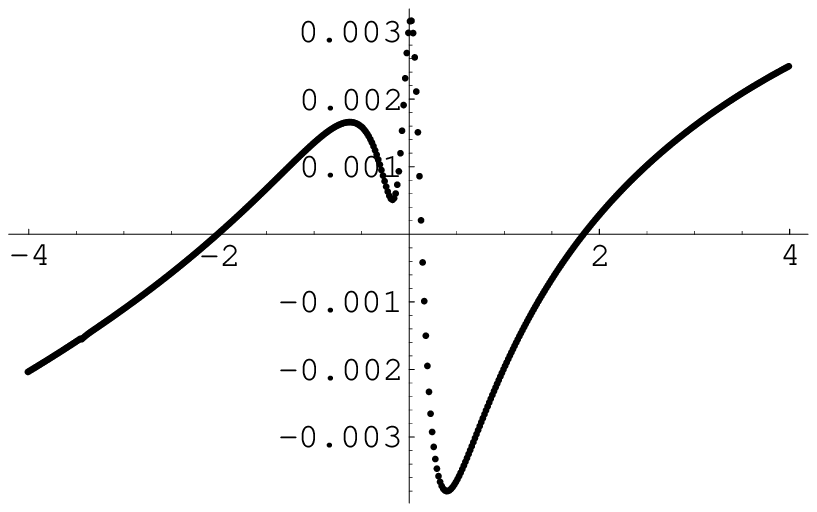,width=2.5in}%,height=75mm}
\end{flushleft} 
\vskip -10pt  
\caption{{\small In the solar barycenter system during Earth flyby, Galileo's kinetic energy 
and the potential energy from the Earth  and next Galileo's total energy 
and the variation of the Jacobi ``constant" ($J$) about -1414.6069 are given, all per unit mass and all plotted in units of (km/s)$^2$ vs. time from periapsis in hours.}    
            \label{gega-allE}}  
\end{figure}

%***********

Figure \ref{gega-allE} first shows, respectively, Galileo's kinetic and potential energy from Earth (per unit mass) during flyby.  
Next in Figure  \ref{gega-allE} is shown (per unit mass) Galileo's total energy and the variation about the fit to the Jacobi constant, -1414.6069 (km/s)$^2$, during flyby. 
For this and the other Earth flybys the zero horizontal reference lines represent the average quoted values of $J$.

The interesting changes are seen in Galileo's kinetic energy and total energy.  As with Pioneer 11, there is a peak in kinetic energy transfer after periapsis, which peak then comes down to the asymptotic value.  This peak is transfered to the total energy.  The peak is smaller in total relative size than Pioneer 11's because of the different scales of the two flybys.   The fit to $J=-1414.6069$ (km/s)$^2$ is better than the accuracy of the calculations. 

With only GEGA1 available to suggest an anomalous Earth flyby, and
with no data to characterize the time history of the anomaly, this 
result was not widely reported. (A second flyby, GEGA2, occurred
on 8 December 1992, but any potential anomaly was masked by the effects of low
altitude drag \cite{giga2}.\footnote{
GEGA2 had an even lower altitude, 303 km, than
GEGA1 and had a flyby
velocity of $v_F=14.0$ km/s. 
This resulted in a gap in DSN coverage of nearly two hours.
In anticipation that an anomalous velocity increase
might occur, a Tracking and Data Relay Satellite (TDRS) was
used to track the Galileo spacecraft during the DSN blackout \cite{giga2}.
This time the data revealed a decrease of $\sim 5.9$ mm/s in
velocity, which could be accounted for reasonably well
by the introduction of an atmospheric drag model and by a single mass point on the Earth's
surface underneath the spacecraft trajectory. 
It was concluded \cite{ant} that if
an anomalous velocity increase comparable to GEGA1 were present, it
would be masked by uncertainties in the fitting model.
})

%********************************** 3.2

\subsection{The NEAR flyby (NEGA)}

On 23 January 1998 the NEAR spacecraft flew by Earth (NEGA)
at a velocity of $v_F=12.7$ km/s and at an altitude in the geocentric system $A=532$ km. An 
analysis of the tracking data revealed that 
an orbital energy increase occurred in the vicinity of closest approach. This was true even though this flyby gave a {\it negative} gravity assist, to reach Eros after the farther-out orbital encounter with Mathilde.  

Here the encounter came from {\it outside} the Earth's orbit, and the peak energy transfer occurred just before periapsis.  Further, the peak was a positive transfer even though the final transfer was negative.  This emphasizes the importance of the vector orientation of the two orbits.    

\begin{figure}[h!]
\begin{flushleft}
% \hskip 30pt
\noindent
\psfig{figure=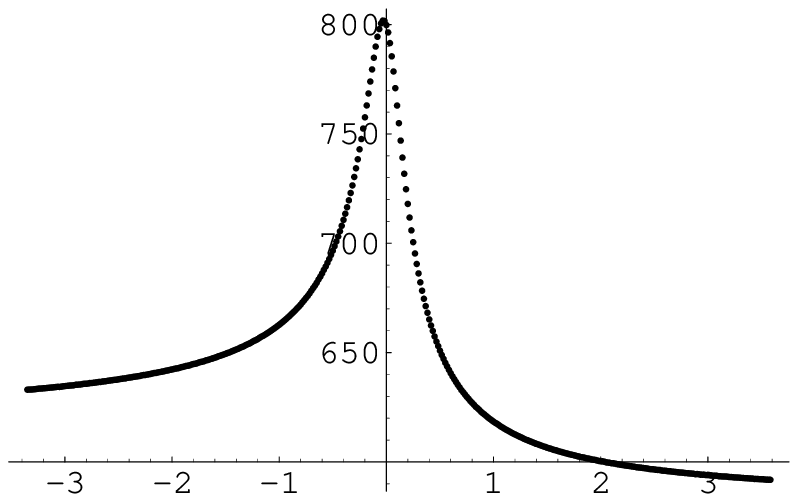,width=2.5in}%,height=75mm}
% \hskip 20pt
\psfig{figure=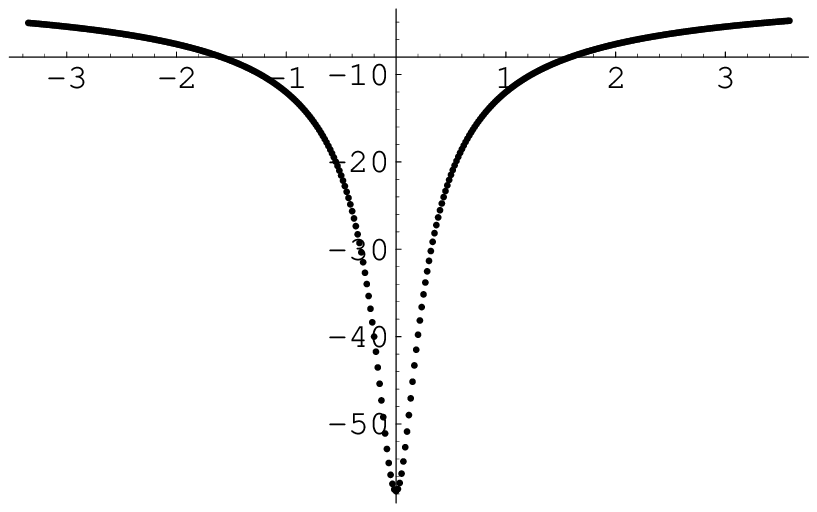,width=2.5in}%,height=75mm}
\end{flushleft} 
\begin{flushleft}
% \hskip 30pt
\noindent
\psfig{figure=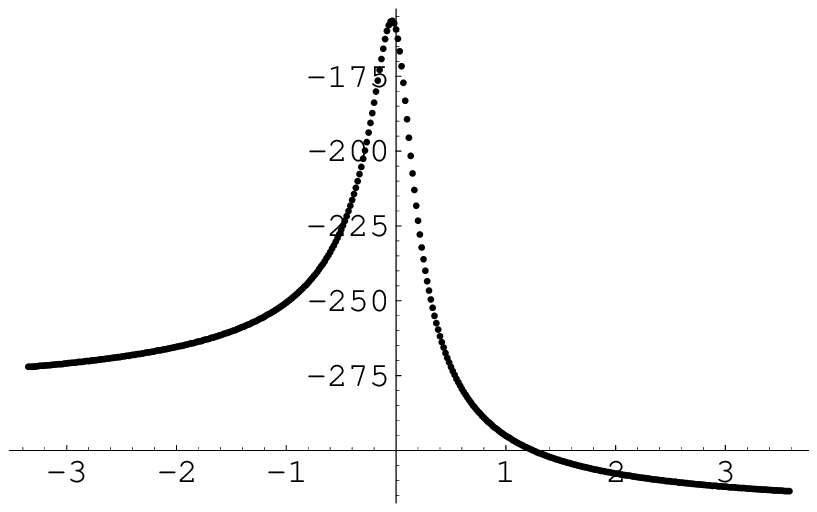,width=2.5in}%,height=75mm}
% \hskip 20pt
\psfig{figure=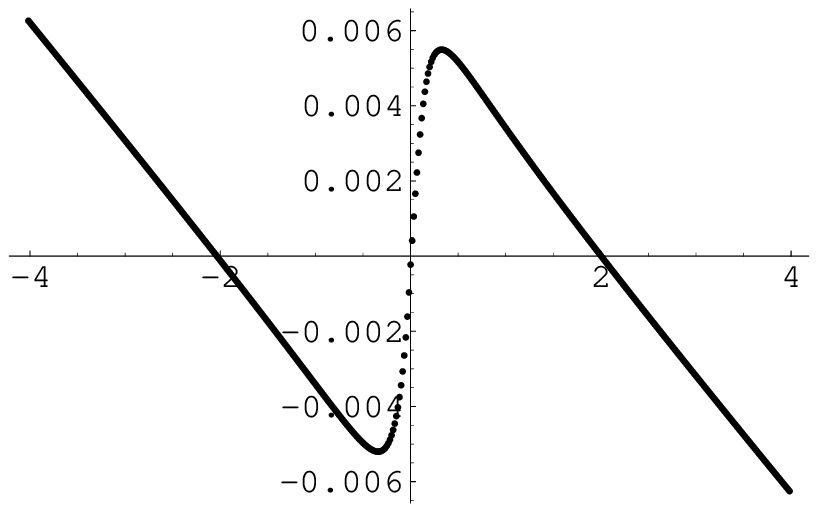,width=2.5in}%,height=75mm}
\end{flushleft} 
\vskip -10pt  
\caption{{\small 
In the Sun's barycenter system, NEAR's kinetic energy 
and the potential energy from the Earth per unit mass during flyby and next NEAR's total energy per unit mass during flyby 
and the fit about the determined  Jacobi ``constant" ($J= -1325.5466$)  per unit mass during flyby are given, all plotted in units of (km/s)$^2$ vs. time from periapsis in hours.}  
            \label{N}}  
\end{figure}

When expressed in terms of the orbital
hyperbolic excess velocity $v_\infty=6.851$ km/s, the
increase for NEAR was \cite{ant} $\Delta v_\infty = (13.46 \pm 0.13)$
mm/s. The equivalent orbital energy increase was $(92.21 \pm 
0.89)$ J/kg.  (See column 3 of Table
\ref{haletable}.)  Figure \ref{N} shows the kinetic energy and the Earth's potential energy on NEAR and NEAR's total energy and the variation about the fit to the Jacobi constant of $J=-1325.5466$ (km/s)$^2$, all per unit mass as a function of time about periapsis. 

%************ Fig 7

%************

%****************************************** 3.3

\subsection{Cassini}

The NEAR result increased interest in the then upcoming Earth flyby by the Cassini spacecraft on 18 August 1999.  
The Cassini flyby altitude was 1171 km and its velocity of $v_F=19.03$ km/s were greater than those
for either GEGA1 or NEGA.  It is the only Earth flyby considered here that was prograde relative to Earth's rotation.  (From a point on the Earth it travels overhead from west to east and hence its inclination is less than $90^\circ$.)  

\begin{figure}[h!]
\begin{flushleft}
% \hskip 30pt
\noindent
\psfig{figure=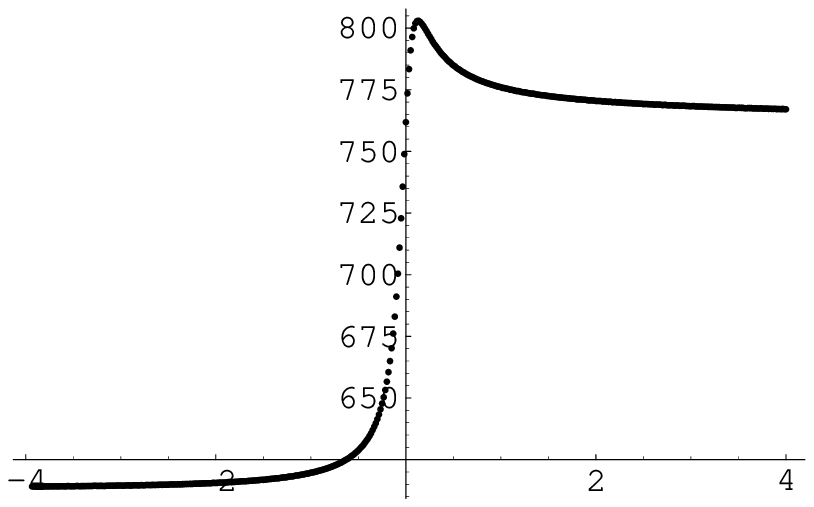,width=2.5in}%,height=75mm}
% \hskip 20pt
\psfig{figure=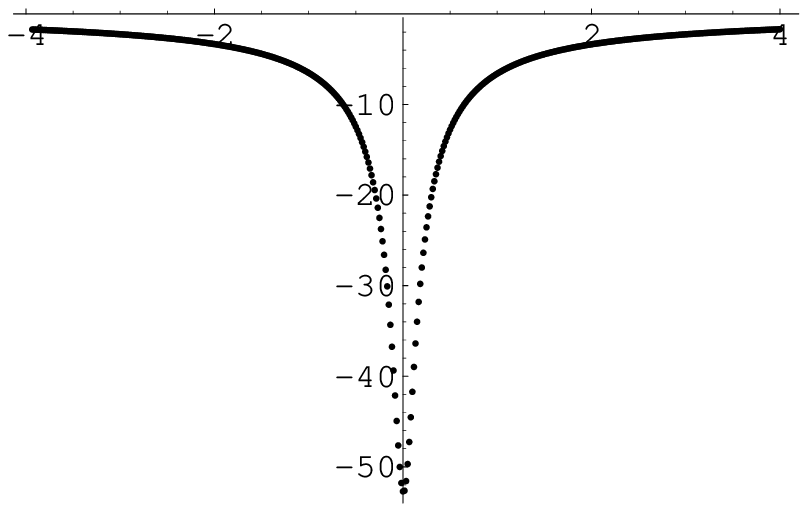,width=2.5in}%,height=75mm}
\end{flushleft} 
\begin{flushleft}
% \hskip 30pt
\noindent
\psfig{figure=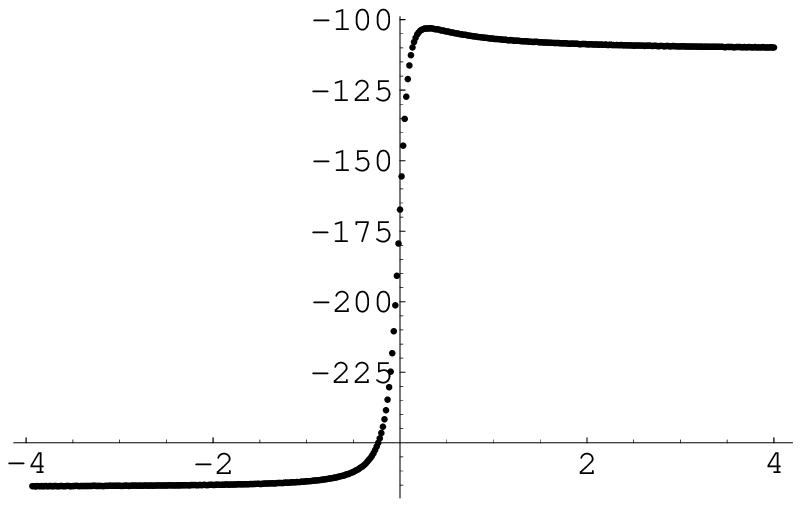,width=2.5in}%,height=75mm}
% \hskip 20pt
\psfig{figure=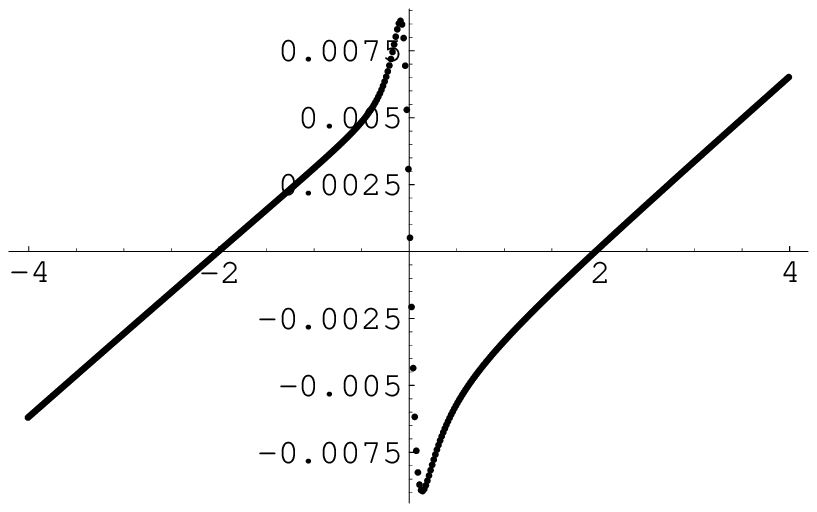,width=2.5in}%,height=75mm}
\end{flushleft} 
\vskip -10pt  
\caption{{\small In the Sun's barycenter system, Cassini's kinetic energy 
and the potential energy from the Earth per unit mass during flyby and next Cassini's total energy per unit mass 
and the fit about the determined  Jacobi ``constant" ($J=-1175.3385 $) per unit mass during flyby are given, all plotted in units of (km/s)$^2$ vs. time from periapsis in hours.}  
            \label{C}}  
\end{figure}

Figure \ref{C} shows the kinetic energy and the Earth's potential energy on Cassini and Cassini's total energy and the fit about the determined Jacobi constant of $J=-1175.3385$ (km/s)$^2$, all per unit mass as a function of time about periapsis. The energy transfer is similar to the Pioneer 11/GEGA cases, but the peaks are much smaller.  

%************ Fig 8

%************

What, then, about any anomalous energy shift?  This time any anomalous orbital energy increase was smaller, not surprising given the larger distance from the Earth \cite{cass}.  
Unfortunately, at the epoch of Cassini periapse, an explicit velocity increment of order 2.1 mm/s occurred as the result of a series of attitude control jet firings \cite{roth}, effectively masking any observation of anomalous energy shift.  
Further analysis of the attitude jet firings will be needed, but for now we take as an upper limit for any anomalous energy shift 10\% of the  velocity increment currently attributed to the attitude jet firings , i.e., -0.2 mm/s. Any significance of the negative sign relative to the prograde motion awaits further explanation.\footnote{  
Soon after, the Stardust spacecraft had an EGA in Jan., 2001.  But there were thruster problems which masked any signal. 
}

%****************************** 3.4

\subsection{Rosetta}

The next Earth flyby was by Rosetta, on 4 March 2005.  It too gave a positive signal of an anomalous energy gain.  
Rosetta was an ESA craft tracked primarily through ESA's European Space Operations Centre (ESOC), and in part by NASA's DSN.  This provided somewhat independent data analysis \cite{rose} for any obtained $\Delta v$.  The results, primarily from Ref. \cite{rose}, are shown in column 5 of Table \ref{haletable}.  There was an anomalous velocity increase of $\Delta v_\infty = (1.82 \pm 0.05)$ mm/s.    

\begin{figure}[h!]
\begin{flushleft}
% \hskip 30pt
\noindent
\psfig{figure=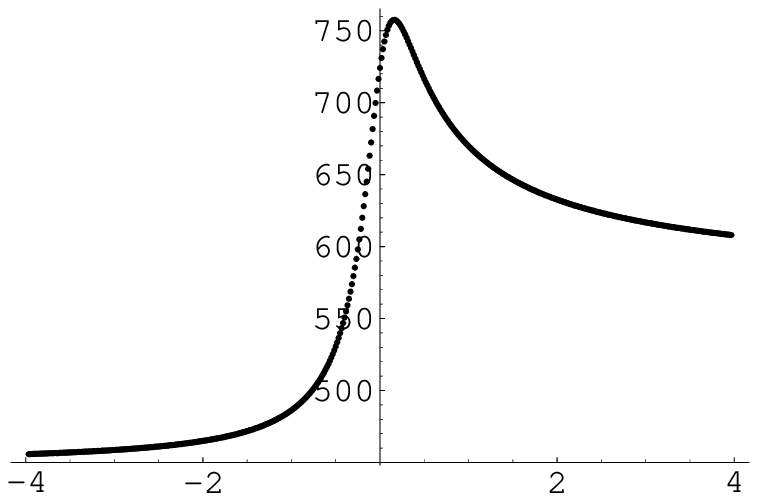,width=2.5in}%,height=75mm}
% \hskip 20pt
\psfig{figure=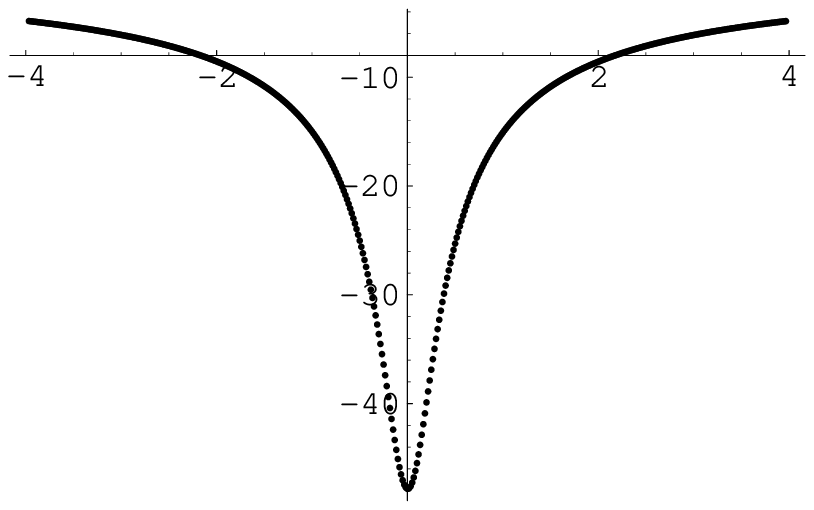,width=2.5in}%,height=75mm}
\end{flushleft} 
\begin{flushleft}
% \hskip 30pt
\noindent
\psfig{figure=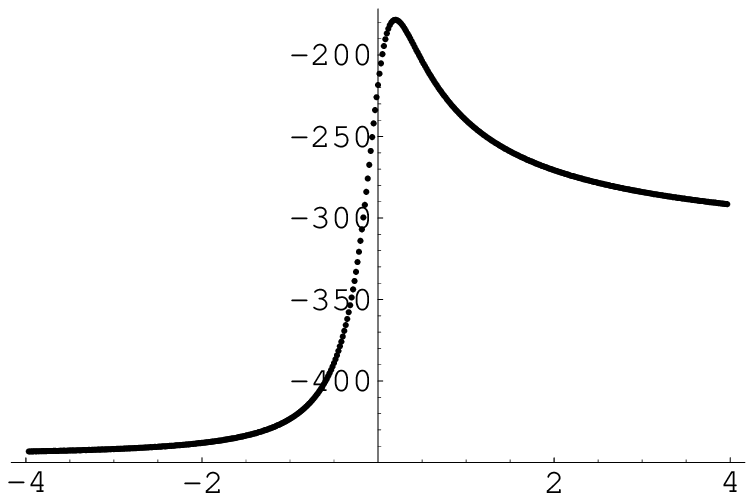,width=2.5in}%,height=75mm}
% \hskip 20pt
\psfig{figure=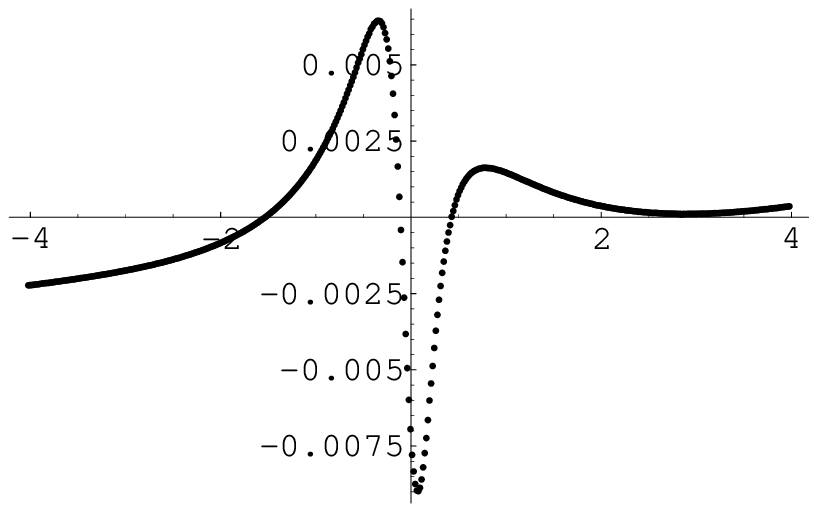,width=2.5in}%,height=75mm}
\end{flushleft} 
\vskip -10pt  
\caption{{\small In the Sun's barycenter system, Rosetta's kinetic energy 
and the potential energy from the Earth per unit mass during flyby and next Rosetta's total energy per unit mass  
and the fit about the determined value of the Jacobi ``constant" ($J=-1274.7442$) per unit mass during flyby are given, all plotted in units of (km/s)$^2$ vs. time from periapsis in hours.}  
            \label{R}}  
\end{figure}

Figure \ref{R} shows the kinetic energy and the Earth's potential energy on Rosetta and Rosetta's total energy and the fit to a Jacobi constant of $J=-1274.7442$ (km/s)$^2$, all per unit mass as a function of time about periapsis.  To calculate this figure we used an Horizons orbit (see footnote \ref{JPLdynamics}) with closest approach at 22:09 UTC on 4 March 2005 and at an altitude of 1954 km.  The energy transfer curves are similar in character to those for Pioneer 11 and GEGA1.

%************ Fig 9

%************

%***************************** 3.5

\subsection{Messenger}

Early analysis results are also in hand from the Messenger craft, which had an EGA on 7 Aug. 2005\footnote{\tt http://messenger.jhuapl.edu/missiondesignLive/Tables} 
at an altitude of 2526 km and a flyby velocity of 10.4 km/s.   (See column 6 of Table \ref{haletable}.)  This, like NEAR,  provided a negative gravity assist, taking energy away to eventually achieve Mercury's orbit.  But the energy transfer curves are very different than NEAR's.   

\begin{figure}[h!]
\begin{flushleft}
% \hskip 30pt
\noindent
\psfig{figure=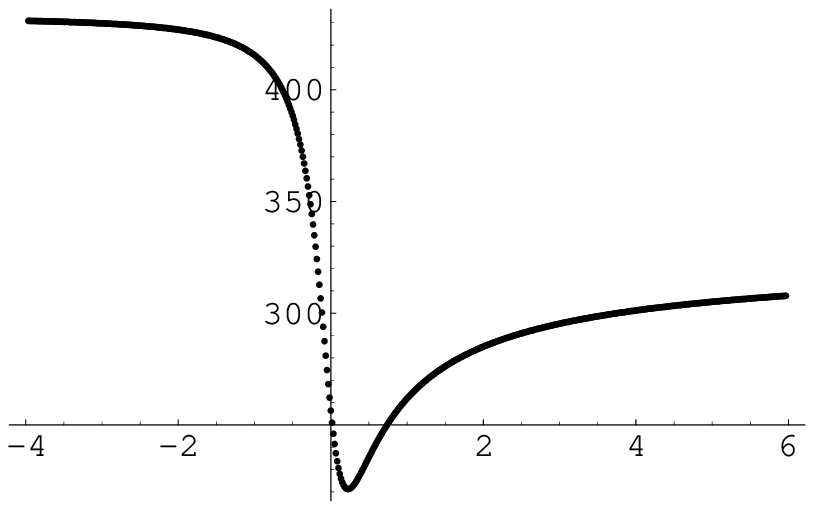,width=2.5in}%,height=75mm}
% \hskip 20pt
\psfig{figure=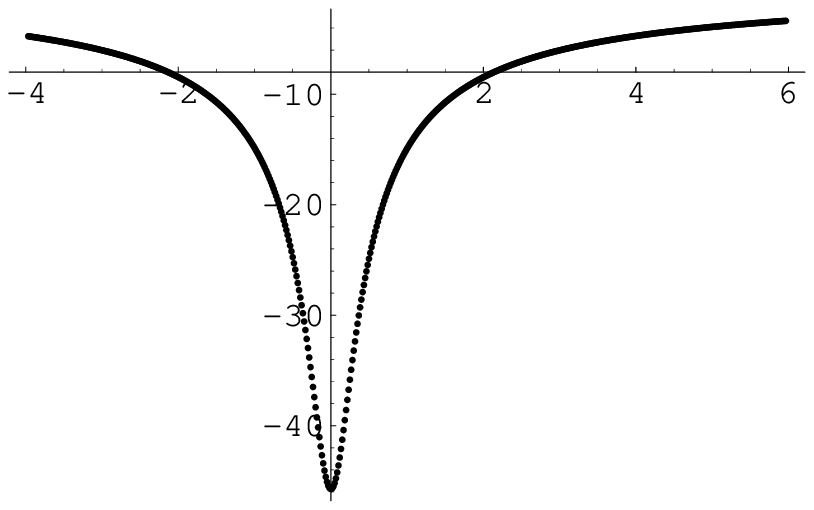,width=2.5in}%,height=75mm}
\end{flushleft} 
\begin{flushleft}
% \hskip 30pt
\noindent
\psfig{figure=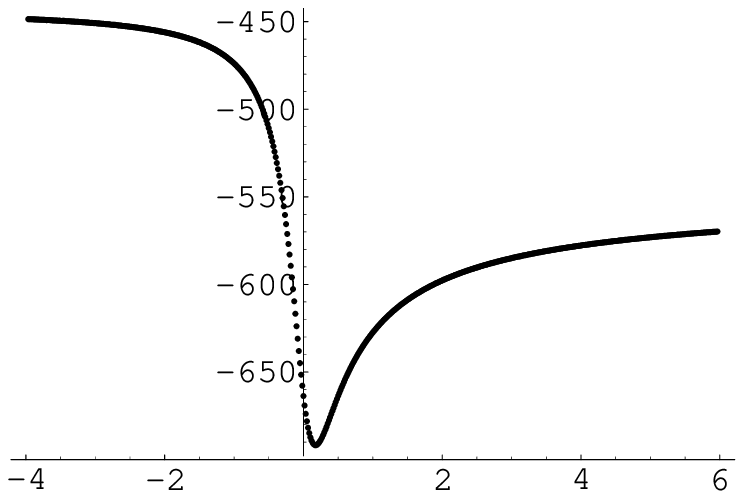,width=2.5in}%,height=75mm}
% \hskip 20pt
\psfig{figure=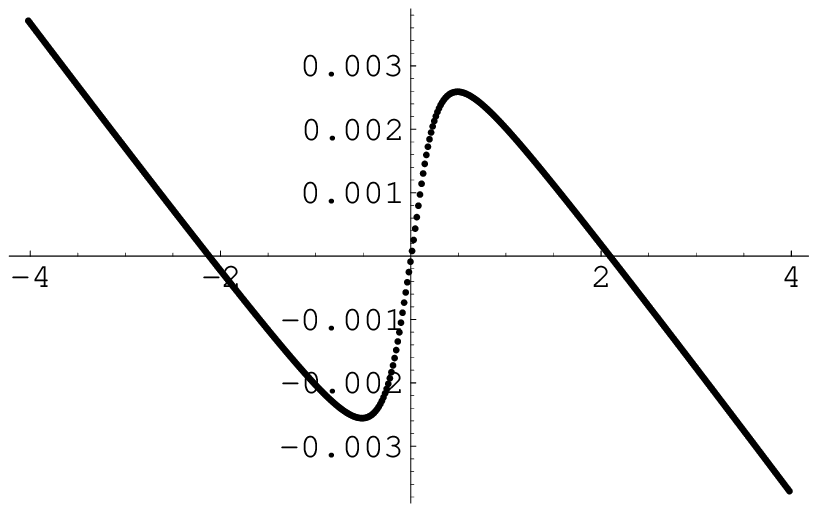,width=2.5in}%,height=75mm}
\end{flushleft} 
\vskip -10pt  
\caption{{\small In the Sun's barycenter system, Messenger's kinetic energy 
and the potential energy from the Earth per unit mass during flyby and next Messenger's total energy per unit mass 
and the fit about the determined value of the Jacobi ``constant" ($J=-1298.5380$) per unit 
mass during flyby are given, all plotted in units of (km/s)$^2$ vs. time from periapsis in hours.}  
            \label{M}}  
\end{figure}

In Figure \ref{M} we plot energy functions for the Messenger flyby.   (Similar to our other calculations, the total potential energy includes contributions from the Sun, the Earth and its quadrapole moment, the Moon, and Jupiter.)  Unlike NEAR, where there was first a positive energy transfer spike, here the figures are more an up-down reflection of the Pioneer 11 and GEGA1 figures.  There is a continuous decrease of energy with time to a negative spike, and then a relaxation to the final transfer energy.

%************ Fig 10

%************

Note that now the kinetic and total energies decrease until just after periapsis and then increase somewhat to their final values.  So this time the energy transfer is more a mirror of the GEGA1 case rather than mimicking the NEAR negative gravity assist. 

The preliminary analysis of the flyby has not, to date, shown any significant evidence for an anomalous energy.   

%******************************* 4

\section{Discussion}
\label{disc}

Having described the dynamical situations of these anomalous flybys, the question arises if there is any insight that might be obtained.   We note again that the best-fit values of the rotational frequencies obtained for the Jacobi constants differ bt 10 to 20 \% from the physical values.  {\it A priori} this could in principle be due to missing mass, a misorientation of the invariable plane, an added force, or a light speed anomaly, all of which could be mapped into one another without better knowledge.    

\begin{figure}[h!]
\begin{flushleft}
% \hskip 15pt
\noindent
\epsfig{figure=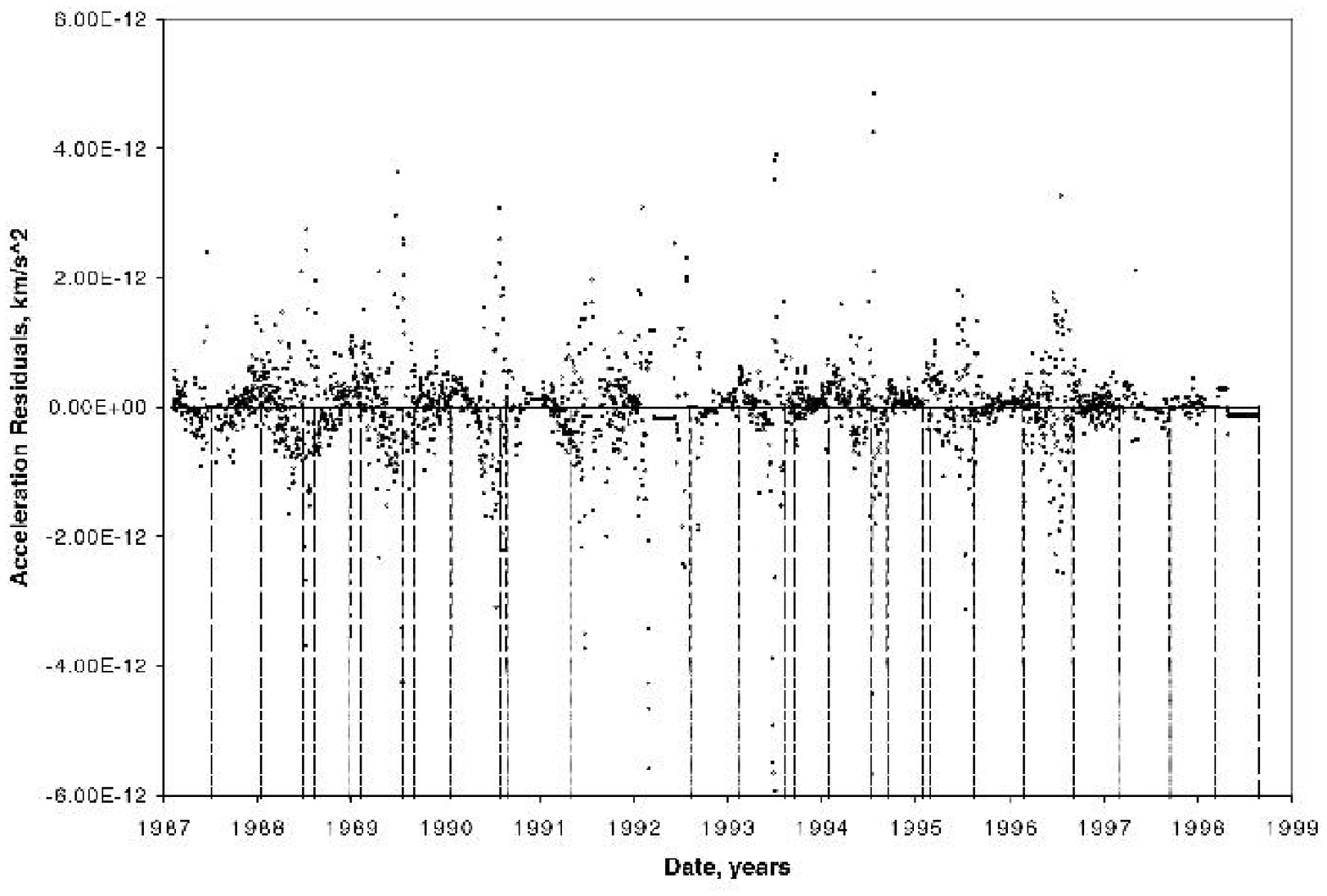,width=2.95in}%,height=85mm}
% \hskip  20pt
\psfig{figure=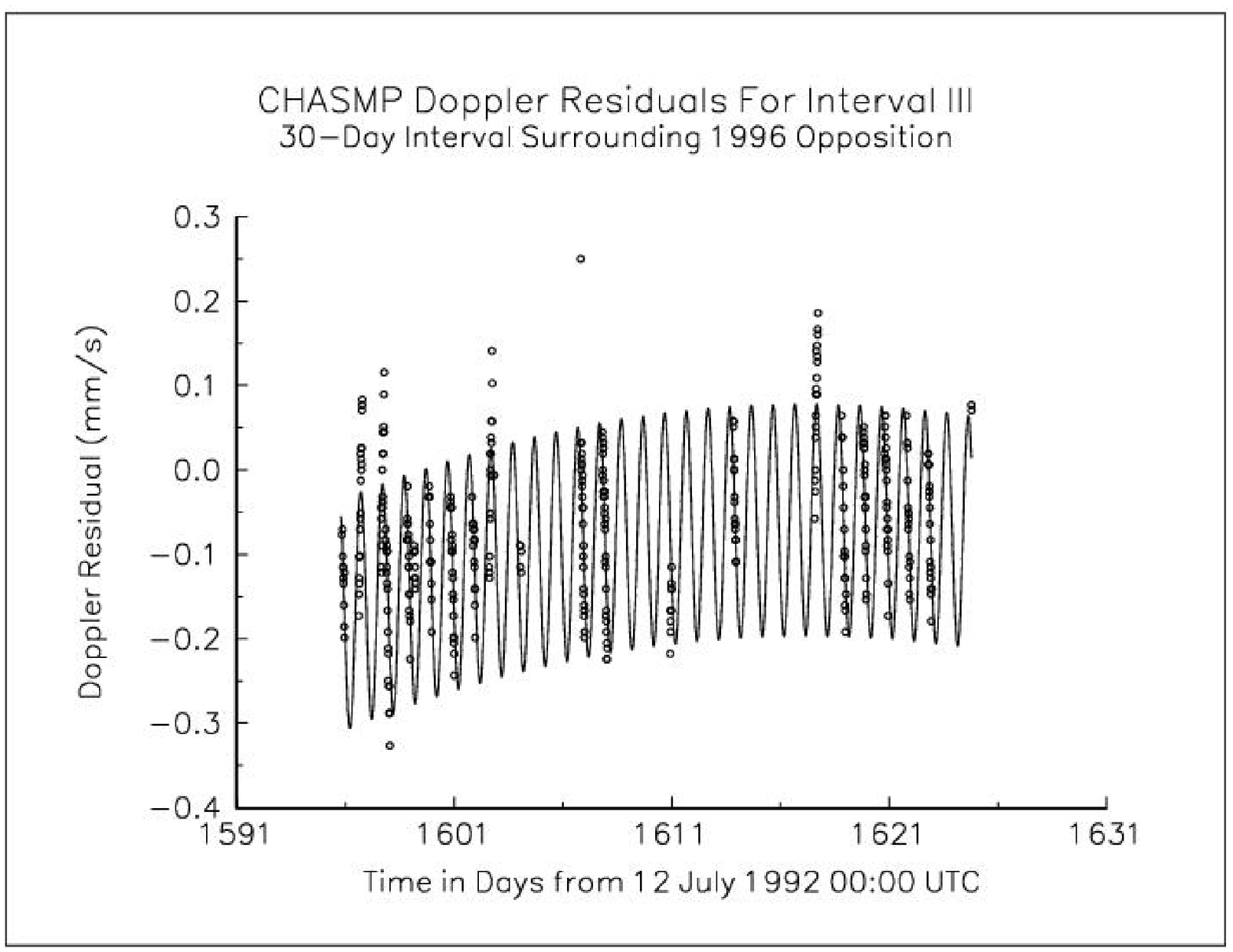,width=2.45in}%,height=4.5in}
\end{flushleft} 
\vskip -10pt  
\caption
{First we give ODP 1-day batch-sequential acceleration
residuals using the Pioneer 10  data set from 1987.0 to 1998.5 \cite{pioprd}.  
Maneuver times are indicated by the vertical dashed lines. The annual term is apparent as an overlay to the constant baseline.  Next we give CHASMP program  acceleration residuals from 23 November 1996 to 23 December 1996 \cite{pioprd}.  
A clear modeling error is represented by the solid diurnal curve.  
(An annual term maximum is also seen as a background.) 
   \label{annual-diur}}
\end{figure}

In this light, it is useful to recall another result from the study of the Pioneer anomaly, 
that there are apparent annual and diurnal terms on top of the constant anomaly \cite{pioprd}.  These are most apparent in clear data sets that came when the Pioneers were at large distances from the Sun.  In Figure \ref{annual-diur} we show examples of this type of signal.   

%*************************************** Fig 11

%************

%******************************

In discussions of the Pioneer anomaly, these annual/diurnal terms should be kept in mind.

Going on to the Earth flybys, 
since the mass distribution of the Earth and the dynamics of
EGAs are supposedly well known, people have started to ask if there is at least some phenomenological  
pattern to the EGA anomalies \cite{awill,cdit}. 

Note that the simplest Newtonian version of the problem is the hyperbolic orbit in the geocentric system: 
\bea
r &=&\frac{a(\epsilon^2 -1)}{1+\epsilon \cos\theta} 
= \frac{r_F(\epsilon + 1)}{1+\epsilon \cos\theta}, ~~~~~~~~
v_\infty = \left[\frac{M_\oplus G}{a}\right]^{1/2}, \\
v_F &=& \left[v_\infty^2 + \frac{2M_\oplus G}{r_F}\right]^{1/2}, ~~~~~~~~~~~~~~
b= \frac{M_\oplus G}{v_\infty^2 \tan(\Theta/2)}.
\eea
In principle three of these parameters, say $v_\infty$, $\epsilon$, and $r_F$, determine the trajectory.  Therefore, one might consider looking for phenomenological patterns using such variables.  

\begin{figure}[h!]
\begin{flushleft}
% \hskip 15pt
\noindent
\psfig{figure=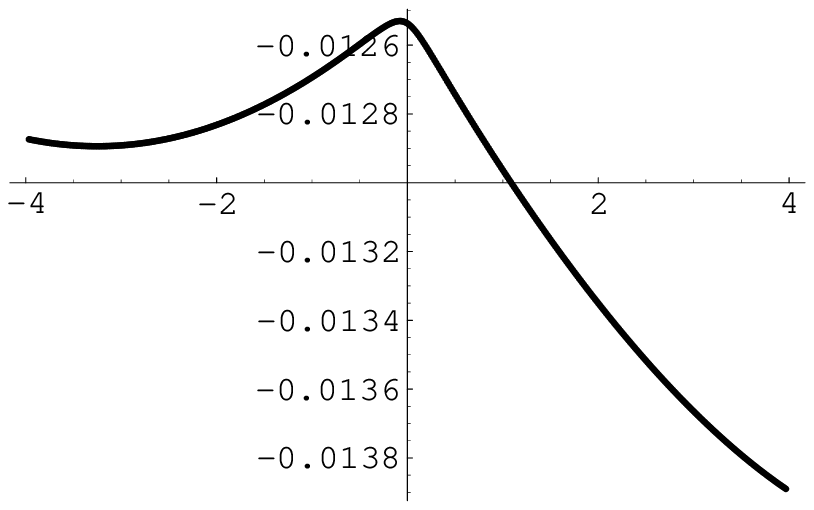,width=1.75in}%,height=75mm}
\hskip 5pt
\psfig{figure=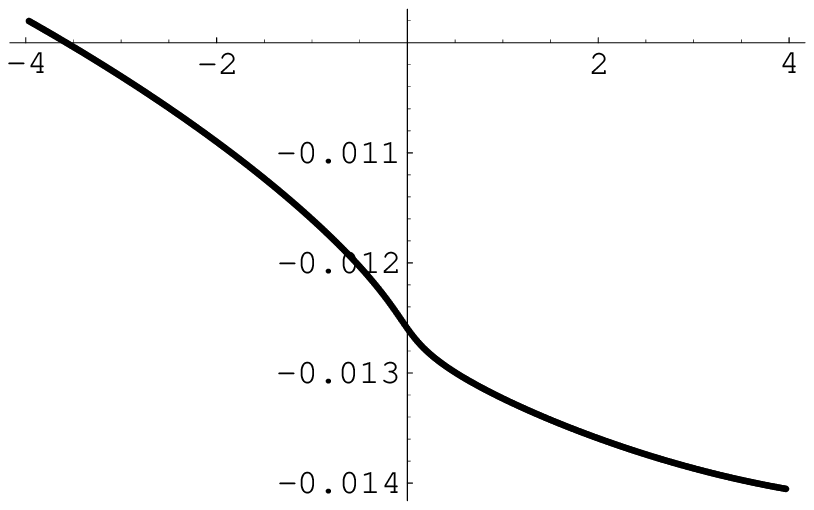,width=1.75in}%,height=75mm}
\hskip 5pt
\psfig{figure=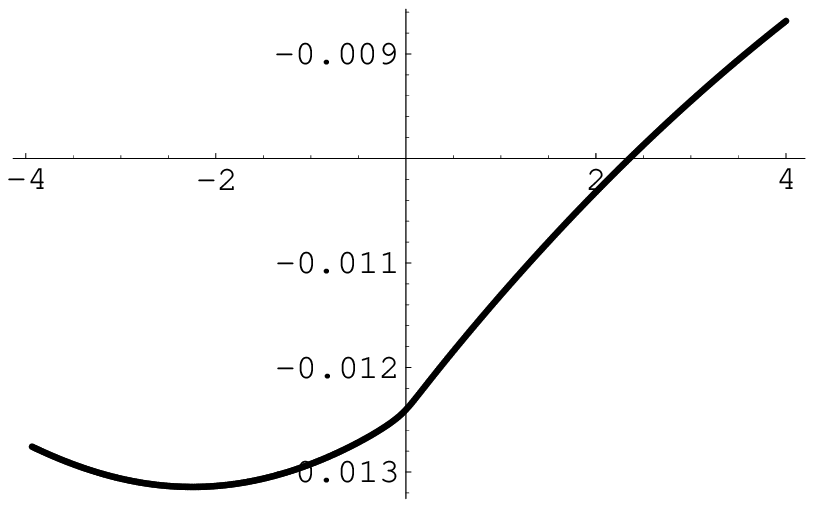,width=1.75in}%,height=75mm}
\end{flushleft}
\begin{flushleft}
\hskip 1in
\noindent
\psfig{figure=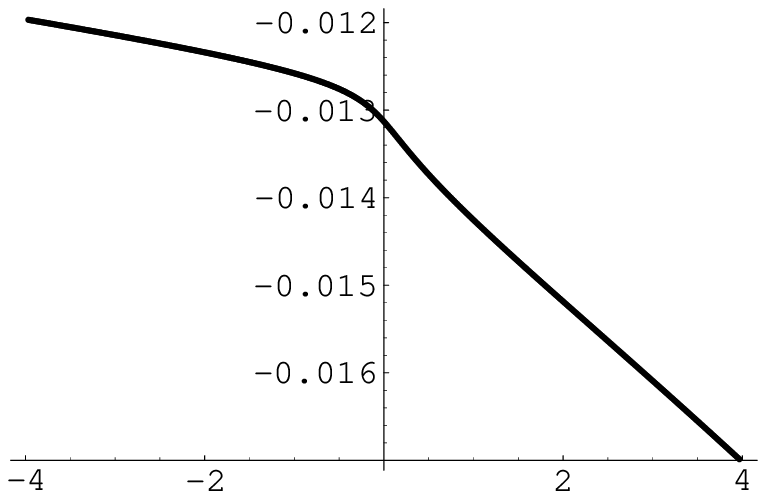,width=1.75in}%,height=75mm}
\hskip 20pt
\psfig{figure=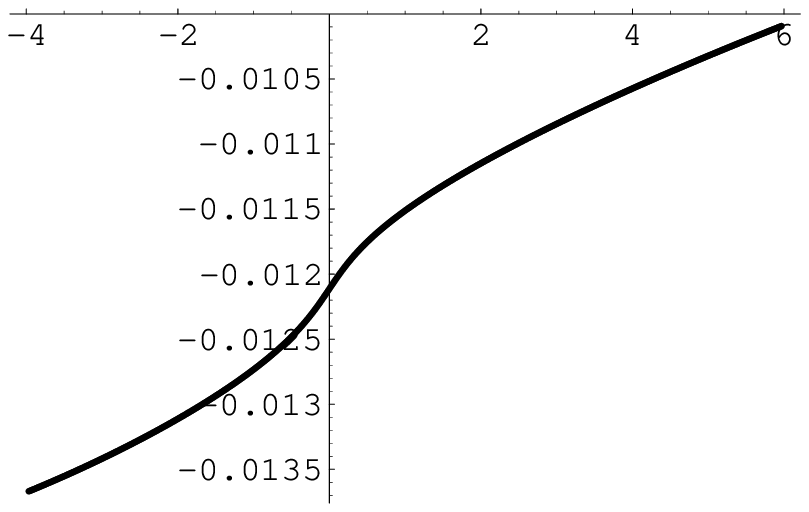,width=1.75in}%,height=75mm}
\end{flushleft} 
\vskip -10pt  
\caption{{\small In the solar barycenter system,  the potential energies (all per unit mass) 
from the Moon during the (in order) GEGA1, NEAR, Cassini, Rosetta, and Messenger flybys in units of (km/s)$^2$ vs. time from periapsis in hours.}  
            \label{moon}}  
\end{figure}

\begin{figure}[h!]
\begin{flushleft}
% \hskip 15pt
\noindent
\psfig{figure=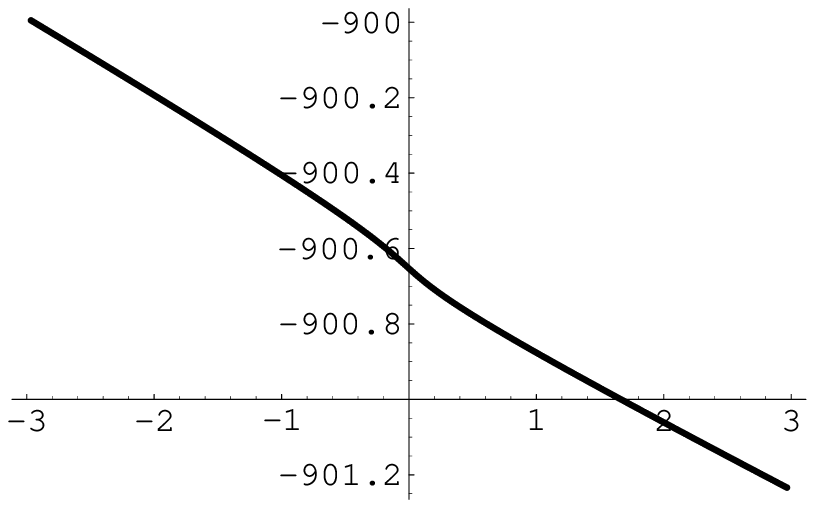,width=1.75in}%,height=75mm}
\hskip 5pt
\psfig{figure=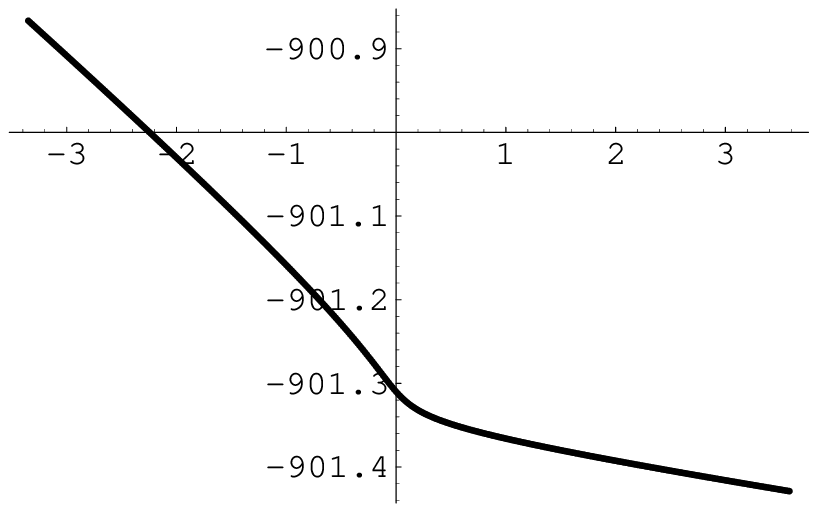,width=1.75in}%,height=75mm}
\hskip 5pt
\psfig{figure=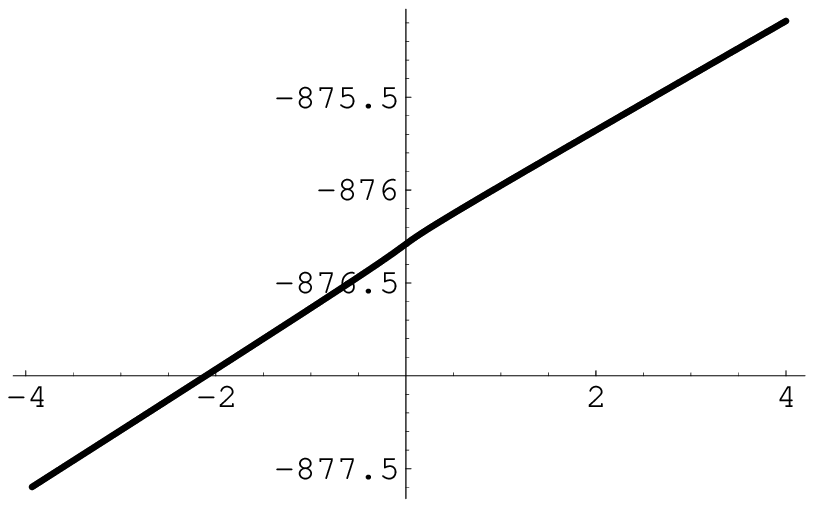,width=1.75in}%,height=75mm}
\end{flushleft}
\begin{flushleft}
\hskip 1in
\noindent
\psfig{figure=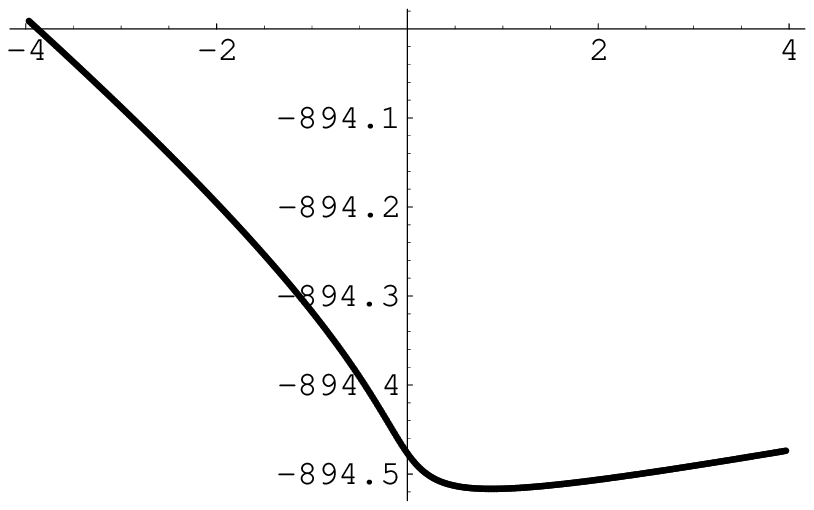,width=1.75in}%,height=75mm}
\hskip 20pt
\psfig{figure=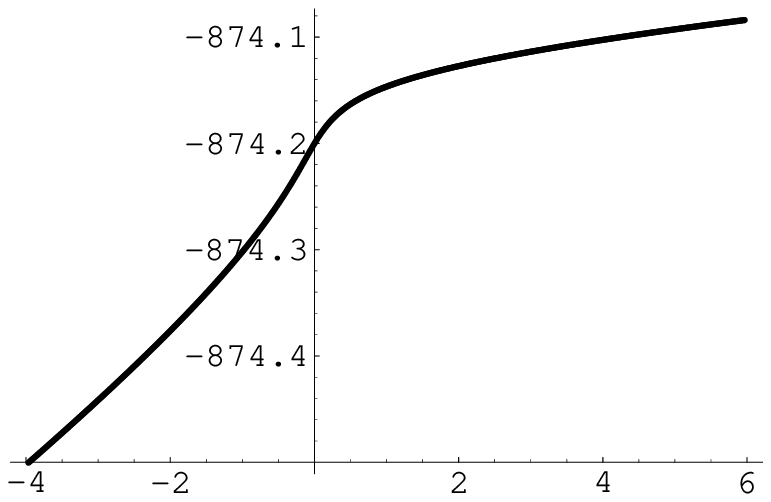,width=1.75in}%,height=75mm}
\end{flushleft} 
\vskip -10pt  
\caption{{\small In the solar barycenter system,  the potential energies (all per unit mass)from the Sun during the (in order) GEGA1, NEAR, Cassini, Rosetta, and Messenger flybys in units of (km/s)$^2$ vs. time from periapsis in hours.}  
            \label{sun}}  
\end{figure}

Another consideration is the largest non-Earth perturbation, which comes from the Moon. (Jupiter has a larger potential contribution but it is the potential gradient (force) that is important.)  In Figure \ref{moon} we show the potential energies from the Moon during the GEGA1, NEAR, Cassini, Rosetta, and Messenger flybys.  The Moon's positional dependence could be important, as it is part of a separate three-body problem (Sun, Earth, Moon), making the actual spacecraft motion closer to a restricted four-body problem.  
Also, in Figure \ref{sun} we show the potential energies from the Sun during the respective Earth flybys.  These are large but do not change much during the flybys.

%************ Fig 12

%************

\vskip .25in

%************ Fig 13

%************

The data points currently available on possible flyby anomalies are limited to the two Pioneer flybys of a giant planet and to the Earth flybys of Table 
\ref{haletable}. Unfortunately, the flybys of large satellites orbiting Jupiter and Saturn are not reliable because of uncertainties in their gravitational fields, as evidenced by Ganymede \cite{anderson2004, palguta}. In addition, the detection of any small flyby anomaly, with respect to the absolute scale of the velocity in the flyby trajectory, requires a spacecraft trajectory relatively free of systematic accelerations. These accelerations can be generated by onboard systems, by non-conservative forces such as atmospheric drag, or by conservative forces from uncertainties in the planet's gravitational field. 

Further, the radio Doppler data must be of high quality and referenced to ground-based atomic frequency standards by means of coherent two-way tracking. This limits the number of candidates for further flyby tests, whether of the Earth or other planets. However, given the importance of the gravity-assist technique for the conservation of rocket fuel, more flybys will occur in the future. Now that possible anomalies have been detected, it is just a matter of waiting for opportunities and then doing the necessary tweaking of the flyby orbit, the onboard systems, and the Doppler tracking in order to optimize the possibility for another measurement of an effect. 

In summary, we have presented a description and brief discussion of the physics of the energy-transfer process that occurs during planetary flybys.
We have also presented a series of intriguing real-world results associated with flybys that belie our current understanding of the underlying physics.  It is hoped that further study, which we encourage, can reconcile this situation.

%**************************************

\section*{Acknowledgements}

We first emphasize that the flyby anomalies discussed here have long been a
concern at JPL, and that we have benefited from the unpublished work of others.
In particular, the authors gratefully acknowledge engineers and scientists in the Guidance, Navigation, and Control Section of the Jet Propulsion Laboratory who have contributed to the analyses of the Earth flyby tracking data and its interpretation.
We also thank R. S. Nerem, Goddard Spaceflight Center,
and X X Shum, University of Texas Center for Space Research, for
verifying that our Earth-flyby Doppler calculation using JPL's
ODP agrees with their geocentric satellite software (Geodyne and UTOPIA). 
Trevor Morley and Bobby Williams kindly gave us information on the Rosetta and Messenger flybys, respectively.  Finally, for many helpful conversations especially on the Pioneer anomaly, but also on the Earth flyby anomalies, we thank Hansjoerg Dittus, Claus L\"ammerzahl, Eunice L. Lau, and Slava G. Turyshev.  
The work of JDA and JKC was performed at the Jet Propulsion Laboratory,
California Institute of Technology, under contract with the
National Aeronautics and Space Administration. MMN acknowledges
support by the US Department of Energy.

%************************** App A

\appendix

\section{Navigation data}
\label{appA}

The navigation results utilized in this paper for Pioneer 10 at Jupiter, Pioneer 11 at Saturn, 
\cite{pioprd}, 
and for the Galileo, NEAR, and Cassini Earth flybys
\cite{ant} 
were produced at JPL/Caltech in Pasadena, California with the Orbit Determination Program (ODP). The results for Messenger were produced at KinetX Corporation in Simi Valley California with their Mirage software 
\cite{williams}, 
a navigational system developed at JPL and therefore profoundly similar to the ODP. 

The results for Rosetta were produced at the European Space Agency’s European Space Operation Center (ESA-ESOC) in Darmstadt Germany. The software is independent of JPL’s ODP, although like ODP it is based on a batch least squares procedure. It is intended primarily to fit Doppler and ranging data from ESA’s 35m New Norcia (NNO) station near Perth Australia, a station of ESA’s Intermediate Frequency Modulation System (IFMS), but it can fit data from DSN tracking stations as well. The Rosetta Radio Science Team, which depends on both IFMS and DSN tracking data, is a collaboration of ESA and NASA scientists 
\cite{patzoldA,patzoldB}.

All these software systems depend on similar fitting models 
\cite{moyer}.
The dynamical models are referred to the J2000 inertial reference frame of the JPL export ephemerides, with Barycentric Dynamical Time (TDB) as the independent variable for the spacecraft’s vector position and velocity. In addition to the Newtonian attraction of the planets and the Moon, including a high degree and order truncated Legendre expansion of the Earth’s gravitational potential, the models include post-Newtonian corrections consistent with the theory of general relativity, models for the reaction forces from solar-radiation pressure on the spacecraft, models for orbital trim maneuvers, and stochastic small forces from gas leaks, nonisotropic thermal emission, and uncoupled control jets. The numerical integration of the equations of motion is referred to the Earth's center of mass for the Earth flybys, although the integration is dynamically consistent with the inertial solar-system barycenter as the dynamical center. The Doppler and ranging data are corrected for transponder delay, refraction in the Earth’s troposphere and ionosphere, and for refraction by interplanetary plasma, the latter being of small concern for the Earth flybys. Corrections are also applied for variability in the length of day (the Earth's rotation) and for polar wandering in the Earth's body-fixed coordinate system.

The data delivered for analysis depends on the ground system, whether DSN or IFMS, and whether closed loop or open loop. For all the Earth flybys, the data is available in the closed loop mode and are extracted at the stations by counting cycles of a sinusoidal signal recorded by digital receivers as a function of time (UTC). The data acquisition and data processing is similar for the DSN and IFMS and the delivered data are compatible. The raw data (cycle count) for the Earth flybys is available in archival tracking data format (ATDF). The first level of data processing is done by the DSN or by IFMS and is delivered as an Orbit Data File (ODF). The ODF consists of samples of Doppler frequency shift, defined as the difference of cycle count at a predetermined Doppler integration time TC divided by TC and referenced to a time variable uplink frequency as recorded by the transmitting station \cite{moyer}. 

The result is called closed-loop two-way Doppler data when the transmitter and receiver are at the same station, and three-way Doppler data when the transmitter and receiver are at different stations. Ranging data is delivered on the ODF at the sample times recorded by the ranging receivers at the station and is in ranging units RU, which can be converted to a round-trip light time in UTC seconds 
\cite{moyer}. 
Any further data processing is done by the data analyst, and it is done with a software system such as JPL’s ODP that accepts an ODF file as input data. In that sense, the data displayed in Table 
\ref{haletable}, 
including the anomalous Earth-flyby results, can be taken as processed tracking data. There is no need to go back to the processed data on the ODF, or the raw data on the ATDF, except for purposes of running independent checks on the data processing that leads to the results of Table 
\ref{haletable}.

%************************************** APP B

\section{Flyby calculations}
\label{appB}

\subsection{Major planet flybys by the Pioneers}
\label{B1} 
The Pioneer calculations used the potentials from the Sun, Jupiter, and Saturn ($GM_S$), the Earth's effect being very small and hence possible to ignore.  
For the Pioneer 11 flyby of Saturn, the specific Newtonian approximation used to calculate $\cal{E}$ was,
\bea
 \cal{E}&=& \frac{1}{2} {v^2} - {\frac{GM_\odot}{r_\odot}}  \nonumber \\ 
&& -{\frac{GM_S}{r_S}} \left[ 1 - {J_2^S} {\left( \frac{R_S}{r_S} \right)^2} {P_2} \left( \sin \phi \right)  
\right. \nonumber \\
&& \left.
- {J_4^S} {\left( \frac{R_S}{r_S} \right)^4} {P_4} \left( \sin \phi \right) - {J_6^S} {\left( \frac{R_S}{r_S} \right)^6} {P_6} \left( \sin \phi \right) \right]
\label{GiantTE}
\eea
where the velocity $v$ is with respect to the solar-system barycenter (SSB) at distance $r_\odot$ with solar constant $GM_\odot$.  

The gravitational constant for the Saturn system is $GM_S$. Saturn's oblateness coefficients are given by $J_2^S$, $J_4^S$, and $J_6^S$ for a reference radius $R_S$, $P_n$ is the Legendre polynomial of degree $n$, and the sine of the latitude $\phi$ is obtained by the scalar product of the unit vector $\mathbf{\hat{k}}$ directed to Saturn's north pole and the unit vector $\mathbf{\hat{r}}_S$ directed from Saturn's barycenter to the spacecraft. The right ascension and declination of the Saturn pole in J2000 coordinates is given by $\alpha_{p}^S$ and $\delta_{p}^S$ so that $\mathbf{\hat{k}} = [\cos \alpha_{p}^S \cos \delta_{p}^S, \sin \alpha_{p}^S \cos \delta_{p}^S, \sin \delta_{p}^S]$. The sixth harmonic coefficient $J_6^S$ is negligible for three-place accuracy in the energy calculation, but it is included. 
The values of the Saturn constants used in the calculation are given in subsection \ref{B3}. 

Jupiter simply adds a constant bias to the energy over the time interval of the flyby, and can be ignored. The flyby is referenced to the barycenter of the Saturn system, rather than the center of the planet, with $r_S$ the distance between the Saturn barycenter and the spacecraft. By using the barycenter, the satellites of the system are accounted for, and the approximation to the energy is more accurate, just so long as the spacecraft does not approach a satellite at close range.

The energy calculation for the Pioneer 10 flyby of Jupiter is formulated exactly the same as in Eq. (\ref{GiantTE}) for Saturn, but with the subscript/superscript S replaced by J.  The values of the Jupiter constants are also given in subsection \ref{B3}.

According to the Horizons trajectory, the closest approach of Pioneer 11 to the Saturn barycenter was 79446 km on 1 September 1979 16:29:23.553 ET or 16:28:33.370 UTC.  The closest approach of Pioneer 10 to the Jupiter barycenter was 202756 km on 4 December 1973 02:25:55.318 ET or 02:25:11.135 UTC, with times referred to the spacecraft. The Pioneer trajectories archived on Horizons possibly could be improved for these two flybys by new fits to the recently-retrieved early Doppler data \cite{edata,stoth}.   
            
%********************************** B2

\subsection{Earth flybys} 
\label{B2}

The energy calculations for the Earth flybys are also based on spacecraft trajectories archived in JPL's Horizons system (see footnote \ref{JPLdynamics}). As such, the accuracy of the spacecraft orbits is limited by whatever trajectory file was delivered to Horizons by a navigation team working for a particular flight project at the time of the flyby. 
Presumably, all the delivered trajectories provide a good fit to the DSN (or ESOC) radio Doppler and range data generated during the flyby. However, even though we can assume the trajectories provide a good fit to the data, the use of the Horizons system to compute orbital energies has its limitations.

A conservative estimate of the error introduced by the Horizons system can be obtained by doing the energy calculations both by the ``observer'' method and also by the``vector'' method and comparing the results. We conclude that the Horizons system introduces an error of no more than 0.1 (km/s)$^2$. We proceed with the energy calculations with the Horizons' data, but restrict the application of ${\cal E}$ to three significant digits.    

In the end, any inconsistencies between the inbound and outbound Earth flyby data have been reconciled by the introduction of a fictitious maneuver at perigee in the direction of the spacecraft motion. For the Galileo, NEAR, and Rosetta flybys, such a maneuver definitely is needed in order to fit all the inbound and outbound data to the noise level with a single trajectory. 

Another potential problem with the energy calculations is that the total orbital energy per unit mass, ${\cal E}$, is computed with the Newtonian approximation in inertial coordinates. However, the spacecraft trajectory and the JPL ephemerides are computed consistently to post-Newtonian order (order $v^2/c^2$).  It is not solely that post-Newtonian terms of order $10^{-8}$ should be added to the energy calculation, although this could be done, but rather that the constants and coordinate positions of the planets and the Moon would be different if the solar-system data were fit with a Newtonian model as opposed to the relativistic model actually used. 

However, if the coordinates are evaluated at the dynamical time TDB of the ephemeris, and the constants of the ephemeris are used in the calculation, the value of ${\cal E}$ should be good dynamically to at least seven significant digits. The Horizons trajectories limit the ${\cal E}$ calculation, not relativity considerations. Consequently, we include only four principal bodies in the Earth flyby calculations; the Earth, along with its oblateness coefficient, the Sun, the Moon, and Jupiter. This assures that the result for ${\cal E}$ is good to three significant digits.

The Newtonian approximation used to calculate ${\cal E}$ for Earth flybys is the following:
\bea 
{\cal E}  &=& \frac{1}{2} {v^2} - {\frac{GM_\odot}{r_\odot}} 
- {\frac{GM_M}{r_M}}    - {\frac{GM_J}{r_J}} \nonumber \\
&&
- {\frac{GM_\oplus}{r_\oplus}} 
\left[ 1 - {J_2^\oplus} {\left( \frac{R_\oplus}{r_\oplus} \right)^2} {P_2} \left( \sin \delta \right)  \right],
\label{TE}
\eea
where the velocity $v$ is with respect to the SSB, and the gravitational constants are $GM_\odot$ for the Sun, $GM_M$ for the Moon, $GM_J$ for the Jupiter system, and $GM_\oplus$ for the Earth. The Earth's oblateness coefficient is $J_2^\oplus$ for a reference radius $R_\oplus$, $P_2$ is the Legendre polynomial of degree two, and we approximate the true geocentric latitude by the declination $\delta$ in inertial coordinates. Because the $J_2^\oplus$ term adds a maximum magnitude of 0.02 (km/s)$^2$ to ${\cal E}$ for the NEAR trajectory, it can be approximated by the declination and still be below our level of assumed systematic error. The spacecraft-body separation distances are $r_\odot$, $r_{M}$, $r_{J}$, and $r_{\oplus}$ for the Sun, Moon, Jupiter's barycenter, and Earth.

%***************************************** B3

\subsection{Numerical values of constants}
\label{B3}

The solar and Saturn constants used in the Pioneer 11 calculation
are the following  (see footnote \ref{JPLdynamics}):\\
\noindent $GM_\odot = 132712440018$ km$^3$/s$^2$  \\
$GM_S = 37940586$ km$^3$/s$^2 = GM_\odot/3497.898$  \\
$R_S = 60330$ km    \\
$J_2^S = 0.0162906$   \\
$J_4^S = -0.000936$  \\
$J_6^S = 0.000086$    \\
$\alpha_{p}^S = 40.58364$ deg \\
$\delta_{p}^S = 83.53804$ deg

The additional Jupiter constants used in the Pioneer 10 calculation are (see footnote \ref{JPLdynamics}).\\
\noindent $GM_{J}$ = 126712766 km$^3$/s$^2= GM_\odot/1047.3486$  \\
$R_{J} = 71492$ km   \\
$J_2^S = 0.01469645$   \\
$J_4^S = -0.00058722$   \\
$J_6^S = 0.00003508$    \\
$\alpha_{p}^J = 268.0567$ deg  \\
$\delta_{p}^J = 64.4953$ deg

The additional constants used in the Earth flyby calculations are (see footnote \ref{JPLdynamics}): \\
\noindent $GM_\oplus = 398600.4415$ km$^3$/s$^2$ \\
$GM_M = 4902.798$ km$^3$/s$^2$  \\
$R_\oplus = 6378.1363$ km \\
$J_2^\oplus = 0.0010826269$

%**************************************

%***************************************

%*********************************************************


\begin{thebibliography}{99}

\bibitem[Anderson 1992]{JPLmemo} Anderson, J. D. 1992 Quarterly Report to NASA/Ames
Research Center,  
{\it Celestial Mechanics Experiment, Pioneer 10/11,} 22 July 1992.
Also see the later quarterly report for the period 1 Oct. 1992 to 
31 Dec. 1992, dated 17 Dec. 1992,  Letter of Agreement ARC/PP017.  
These contain the present 
Figure \ref{fig:correlation}.

\bibitem[Anderson 1997]{jdaEGA}  Anderson, J. D. 1997 ``Gravity-Assist Navigation," in: J. H. Shirley and R. W. Fairbridge, eds., {\it Encyclopedia of Planetary Sciences} (Chapman \& Hall, New York, 1997), pp. 287-289. 

\bibitem[Anderson 1998]{pioprl}  Anderson, J. D., P. A. Laing, E. L. Lau, A. S. Liu,
M. M. Nieto,  and S. G. Turyshev 1998  
``Indication, from Pioneer 10/11, Galileo, and Ulysses data, of an 
Apparent Anomalous, Weak, Long-Range Acceleration," 
{ Phys. Rev. Lett.} {\bf 81}, 2858-2861.
ArXiv gr-qc/9808081.

\bibitem[Anderson 2001]{awill}  Anderson J. D. and J. G. Williams 2001 
``Long-Range tests of the Equivalence Principle,"
Class. Quant. Grav. {\bf 18}, 2447-2456.

\bibitem[Anderson 2002]{pioprd} J. D. Anderson, J. D., P. A. Laing, E. L. Lau, A. S. Liu,
M. M. Nieto, and S. G. Turyshev 2002
``Study of the Anomalous Acceleration of  Pioneer 10 and 11,"  
Phys. Rev. D. {\bf 65} 082004/1-50.   
ArXiv gr-qc/0104064.

\bibitem[anderson 2004]{anderson2004}
Anderson, J. D., G. Schubert, R. A. Jacobson, E. L. Lau, W. B. Moore, J. L. Palguta 2004 "Discovery of Mass Anomalies on Ganymede," Science {\bf 305}, 989-991.

\bibitem[Antreasian 1998]{ant} Antreasian, P. G. and J. R.  Guinn 1998  
``Investigations into the Unexpected Delta-V Increases during the Earth Gravity Assists of Galileo and NEAR," 
paper no. 98 - 4287
presented at the AIAA/AAS Astrodynamics Specialist Conference and
Exhibit (Boston, August 10 - 12, 1998), 13 pp.  Available at 
{\tt  http://www2.aiaa.org/citations/mp-search.cfm}

\bibitem[Campbell 1991]{gega1} Campbell, J. K. 1991 ``Request for Check of Telemetry Acquired During Earth Flyby Period," Jet Propulsion Laboratory Interoffice Memorandum 314.3-975 to R. W. Koesis, 28 May 1991. 

\bibitem[Danby 1988]{danby} Danby, J. M. A. 1988 {\it Fundamentals of Celestial Mechanics}, Second Edition (Willmann-Bell, Richmond, VA).   (See Chap. 8.)  

\bibitem[D'Amario 1992]{damar} D'Amario, L. A., L. E. Bright, and A. A. Wolf 1992 
``Galileo Trajectory Design," 
in {\it The Galileo Mission}, ed. C. T. Russell,  Space Sci. Rev. {\bf 60}, 
23-78.

\bibitem[Delva 1979]{delva1979} Delva, M., and R. Dvorak 1979 
``Eine Reihenentwicklung des Jacobi Integrals im elliptischen Dreik\"orperproblem,"
Astron. Astrophys. {\bf 77}, 252-254.

\bibitem[Edwards 1994]{giga2}  Edwards, C., J. Anderson, P. Beyer, S. Bhaskaran, J. Border,
S. DiNardo, W. Folkner, R. Haw, S. Nandi, F. Nicholson, T. 
Ottenhoff,  and S. Stephens 1994  
``Tracking Galileo at Earth-2 Perigee Using the Tracking and Delay Relay Satellite System ," 
Advances in the Astronautical Sciences {\bf 85}, 1609-1620;
{\it Proceedings of the AAS/AIAA Astrodynamics Conference, Aug 16-19 1993, Part 2; Astrodynamics 1993}, eds. A. K. Misra, V. J. Modi, R. Holdaway, and P. M. Bainum (Univelt, San Diego).

\bibitem[Flandro 1963]{flandroref}   Flandro, G. A. 1966
``Fast Reconnaissance Missions to the Outer Solar System Utilizing Energy Derived from the Gravitational Field of Jupiter,"
Astronautica Acta {\bf 12}, 329-337.

\bibitem[Guman 2002]{cass}  Guman, M. D., D. C. Roth, R. Ionasescu, T. D. Goodson, A. H. Taylor, and J. B. Jones 2000  
``Cassini Orbit Determination from First Venus Flyby to Earth Flyby,"
Advances in the Astronautical Sciences {\bf 105}, 1053-1072 (2000);
{\it Proceedings of the AAS/AIAA Space Flight Mechanics Meeting; January 23-26, 2000}, eds.  C. A. Kluever, B. Neta, C. D. Hall, and J. M. Hanson (Univelt, San Diego).

\bibitem[Hills 1970]{hills} Hills, J. G. 1070 
``Dynamic Relaxation of Planetary Systems and Bode's Law,"
Nature {\bf 225}, 840-842 (1970).

\bibitem[Hohmann 1925]{hohmann}  Hohmann, W. 1925 {\it Die Errichbarkeit der Himmelsk\"orper:  Untersuchungen \"uber das Raumfahrtproblem}  (R. Oldenbourg, Munich).  English translation: {\it The Availability of Heavenly Bodies}, NASA report NASA-TT-F-44 (1960).  

\bibitem[L\"ammerzahl 2006]{cdit}  L\"ammerzahl, C., O. Preuss, and H. Dittus (2006)
``Is the Physics within the Solar System Really Understood?"
ArXiv gr-qc/0604052.  

\bibitem[Minovich 1963]{min}  Minovitch, M. A. 1963
``The Determination and Characteristics of Ballistic Interplanetary Trajectories under the Influence of Multiple Planetary Attractions," 
Report Number: NASA-CR-53033; JPL-TR-32-464.

\bibitem[Morley 2006]{rose} Morley T. and F. Budnik 2006  
``Rosetta Navigation at its First Earth-Swingby," 19th 
Int. Symp. on Space Flight Dynamics 2006, paper ISTS 2006-d-52.
% Kanazawa, Japan, June 6-9
% June/8th (Thu) AM(2) Session d-10 at Room B Deep Space Navigation
 

\bibitem[Moulton 1970]{moulton}  Moulton, F. R. 1970 {\it An Introduction to Celestial Mechanics}, 2nd Rev. Ed. (Dover, Nes York).  See Chap. VIII, pp. 277 ff., esp. p. 319.

\bibitem[Moyer 2000]{moyer}  Moyer, T. D. 2000 ``Formulation for Observed and Computed Values of Deep Space Network Data Types for Navigation," JPL Publication 00-7. 

\bibitem[Nieto 2005]{edata}  Nieto, M. M.  and J. D. Anderson 2005 
``Using Early Data to Illuminate the Pioneer Anomaly," 
Class. Quant. Grav. {\bf 22}, 5343-5354.
ArXiv gr-qc/0507052.

\bibitem[Palguta 2006]{palguta}
Palguta, J., J. D. Anderson, G. Schubert, W. B. Moore 2006 ``Mass Anomalies on Ganymede,"  Icarus {\bf 180}, 428-441.

\bibitem[P\"atzold 2001a]{patzoldA} P\"atzold, M., A. Wennmacher, B. H\"ausler, W. Eidel, T. Morley, N. Thomas, and J. D. Anderson 2001 ``Mass and Density Determinations of 140 Siwa and 4979 Otawara as expected from the Rosetta flybys,"  Astron. Astrophys. {\bf 370}, 1122-1127.

\bibitem[P\"atzold 2001b]{patzoldB}  
P\"atzold, M., B. H\"ausler, A. Wennmacher, K. Aksnes, J. D. Anderson, S. W. Asmar, J. P. Barriot, H. Boehnhardt, W. Eidel, F. M. Neubauer, O. Olsen, J. Schmitt, J. Schwinger, and N. Thomas 2001 ``Gravity Field Determination of a Comet Nucleus: Rosetta at P/Wirtanen,"
Aston. Astrophys. {\bf 375}, 651-660. 

\bibitem[Roth 2006]{roth} Roth, D. (2006) private communication, on results of Roth and T. Barber.

\bibitem[Russell 1992]{russell} Russell, C. T. 1992 {\it The Galileo Mission} (Kluwer, Boston). See especially pp. 24-39, 50-52 and 76-78.

\bibitem[Turyshev 2006]{stoth} Turyshev, S. G., V. T. Toth, L. R. Kellogg, E.. L. Lau, and K. J. Lee 2006  
``The Study of the Pioneer Anomaly: New Data and Objectives for New Investigation,"
Int. J. Mod. Phys. D {\bf 15}, 1-56.  ArXiv gr-qc/0512121.

\bibitem[Van Allen 2003] {VA} Van Allen, J. A. 2003
``Gravitational Assist in Celestial Mechanics - a Tutorial," 
Am. J. Phys. {\bf 71}, 448-451 (2003). 

\bibitem[Wiesel 1989]{wiesel}  Wiesel, W. E. 1989 {\it Spaceflight Dynamics} (McGraw - Hill, New
York).

\bibitem[Williams 2005]{williams} Williams, B., A. Taylor, E. Carranza, J. Miller, D. Stanbridge, B. Page, D. Cotter, L. Efron, R. Farquhar, J. McAdams, and D. Dunham 2005   
``Early Navigation Results for NASA's MESSENGER Mission to Mercury," 
Advanc. Astronaut. Sci. {\bf 120}, 1233-1250;
15th AAS/AIAA Space Flight Mechanics Conference, Paper AAS 05-176, (Copper Mountain, CO, January 23-27, 2005).


\end{thebibliography}
\end{document}